\documentclass{aa}  

\usepackage{graphicx}
\usepackage{txfonts}
\usepackage{natbib}
\usepackage{amsmath} 
\usepackage{amssymb}
\usepackage{mathtools}
\usepackage{xcolor}
\usepackage{multirow}
\usepackage{array}
\usepackage{makecell}
\usepackage{siunitx}
\usepackage{ulem}
\usepackage{graphicx}
\usepackage{subcaption}
\usepackage{verbatim}
\usepackage{hyperref}
\hypersetup{
    breaklinks,
    colorlinks,
    citecolor=blue,
    linkcolor=blue,
    urlcolor=blue,
    linktoc = page,
    pdftitle = {GGSL identification with DL},
    pdfkeywords = {ML, DL, ggsl},
    pdfauthor = {Giuseppe Angora},
    pdfcreator = {\LaTeX}
}
\hypersetup{pdfauthor={Giuseppe Angora}}
\newcommand{\sersic}{S\'ersic }

\begin{document}

   \title{Searching for galaxy-scale strong-lenses in galaxy clusters with deep networks - I: methodology and network performance   }


    \author{G.~Angora\inst{\ref{unife}, \ref{oacn}}\fnmsep\thanks{\email{gius.angora@gmail.com}},
           P.~Rosati\inst{\ref{unife},\ref{oabo},\ref{infnfe}} \and
           M.~Meneghetti\inst{\ref{oabo}} \and
           M.~Brescia\inst{\ref{oacn},\ref{unina}} \and
           A.~Mercurio\inst{\ref{oacn},\ref{unisa}} \and
           C.~Grillo\inst{\ref{unimi},\ref{iasfmi}},
           P.~Bergamini\inst{\ref{unimi},\ref{oabo}} \and
           A.~Acebron \inst{\ref{unimi},\ref{iasfmi}} \and
           G.~Caminha\inst{\ref{tum},\ref{max_plank}} \and        
           M.~Nonino\inst{\ref{oats}} \and        
           L.~Tortorelli\inst{\ref{uni_munich}}\and
           L.~Bazzanini\inst{\ref{unife}, \ref{oabo}} \and
           E.~Vanzella\inst{\ref{oabo}}
          }

   \institute{Department of Physics and Earth Science of the University of Ferrara, Via Saragat 1, I-44122 Ferrara, Italy\label{unife}
   \and
   INAF -- Astronomical Observatory of Capodimonte, Salita Moiariello 16, I-80131 Napoli, Italy\label{oacn}
   \and
   INAF -- OAS, Astronomical Observatory of Bologna, via Gobetti 93/3, 40129 Bologna, Italy\label{oabo}
   \and
   Department of Physics ``E. Pancini'', University of Napoli ``Federico II'', Via Cinthia 21, 80126, Napoli, Italy \label{unina}
   \and  
   Department of Physics of the University of Milano, via Celoria 16, I-20133 Milano, Italy\label{unimi}
   \and
   INAF -- IASF Milano, via A. Corti 12, I-20133 Milano, Italy \label{iasfmi}
   \and
    Technische Universit\"at M\"unchen, Physik-Department, James-Franck Str. 1, 85741 Garching, Germany \label{tum}
    \and
    Max-Planck-Institut f\"ur Astrophysik, Karl-Schwarzschild-Str. 1, D-85748 Garching, Germany \label{max_plank}
   \and
   INAF -- Astronomical Observatory of Trieste, via G. B. Tiepolo 11, I-34131, Trieste, Italy\label{oats}
   \and
   INFN, Sezione di Ferrara, Via Saragat 1, 44122 Ferrara, Italy\label{infnfe}
   \and
   Department of Physics of the University of Salerno - Via Giovanni Paolo II, 132, 84084, Fisciano (SA), Italy \label{unisa}
   \and
   University Observatory, Faculty of Physics, Ludwig-Maximilians-Universität München, Scheinerstr. 1, 81679 Munich, Germany \label{uni_munich}
   }

   \date{Received xxx; accepted xxx}

 
  \abstract
   {Galaxy-scale strong lenses in galaxy clusters provide a unique tool to investigate their inner mass distribution and the sub-halo density profiles in the low-mass regime, which can be compared with the predictions from $\Lambda$CDM cosmological simulations.
   We search for galaxy-galaxy strong-lensing systems in the HST multi-band imaging of galaxy cluster cores from the CLASH and Hubble Frontier Fields programs by exploring the classification capabilities of deep learning techniques.
  Convolutional neural networks are trained utilising highly-realistic simulations of galaxy-scale strong lenses injected into the HST cluster fields around cluster members. To this aim, we take advantage of extensive spectroscopic information on member galaxies in 16 clusters and the accurate knowledge of the deflection fields in half of these from high-precision strong lensing models. Using observationally-based distributions, we sample the magnitudes (down to $F814W = 29$AB), redshifts and sizes of the background galaxy population. By placing these sources within the secondary caustics associated with the cluster galaxies, we build a sample of $\sim\! 3000$ galaxy-galaxy strong lenses which preserve the full complexity of real multi-colour data and produce a wide diversity of strong lensing configurations.
   We study two deep learning networks processing a large sample of image cutouts in three HST/ACS bands, and we quantify their classification performance using several standard metrics. We find that both networks achieve a very good trade-off between purity and completeness ($85\%$--$95\%$), as well as a good stability with fluctuations within $2\%$--$4\%$. We characterise the limited number of false negatives and false positives in terms of the physical properties of the background sources (magnitudes, colours, redshifts and effective radii) and cluster members (Einstein radii and morphology). We also demonstrate the neural networks' high degree of generalisation by applying our method to HST observations of 12 clusters with previously known galaxy-scale lensing systems.
   }


   \titlerunning{Searching for galaxy-scale strong-lenses with deep networks}
   \authorrunning{G. Angora, P. Rosati, M. Meneghetti, M. Brescia, et al.}

   \maketitle
%

\section{Introduction}

Strong gravitational lensing is a powerful tool for studying the mass distribution of galaxies and galaxy clusters and for testing cosmological models. 
Over the last decades, strong lensing has been exploited, for example, to analyse galaxy structures and study their evolution \citep[e.g.][]{Treu2002, Auger2010, Sonnenfeld2013}, to measure the value of the Hubble constant using time delay measurements \citep[e.g.][]{Suyu2017,Suyu2020, grillo2018, Millon2020, Moresco2022}, to constrain the dark energy equation of state \citep[e.g.][]{Jullo2010, Cao2012, Collett2014, Caminha2022}, and to estimate the dark matter fraction in massive early-type galaxies \citep[e.g.][]{Grillo2010, Tortora2010, Sonnenfeld2015}. On cluster-scales, strong lensing models allow for the study of the inner total mass distribution of clusters by exploiting an increasing number of multiple images of background sources  \citep[e.g. ][]
{caminha2017b, caminha2019, Acebron2018, bergamini2019, Bergamini_m0416, lagattuta2019, lagattuta2022}.
In addition, the strong lensing magnification enables clusters to be used as cosmic telescopes to study explore the intrinsic properties of high-redshift faint (lensed) sources, otherwise undetectable \citep[e.g.][]{Swinbank2009, Richard2011, Vanzella2020, Vanzella2021}. 

Recently, by utilising cluster mass maps from high-precision strong-lensing models, \cite{meneghetti2020, Meneghetti2022} reported an excess of galaxy-galaxy strong lensing (GGSL) events in galaxy clusters compared with the expectations from the $\Lambda$CDM structure formation model. This has opened a debate on whether such an excess could be due to limitations in cosmological simulations (e.g. in the mass resolution or in the treatment of baryonic physics) or to more fundamental aspects related to the properties of dark-matter (\citealt{Meneghetti2022} and references therein).

We focus this study on the search for GGSL systems embedded in galaxy cluster halos. In this environment, the probability of GGSLs is generally higher than in the field, for a given lens mass, owing to the contribution of the cluster-scale lensing effect. Traditionally, GGSLs are identified through the visual inspection of candidates selected with spectroscopic or photometric criteria \citep[e.g.][]{Lefevre1988, Jackson2008, Sygnet2010, Pawase2014}. However, this will not be a viable method with the upcoming data-intensive surveys based on the next-generation facilities, such as the European Space Agency (ESA) Euclid satellite \citep{Laureijs2011} and the Vera Rubin Observatory \citep{LSST2019}, which are expected to find tens of thousands of galaxy clusters and $\sim10^5$ GGSLs \citep{LSST2012, Euclid2019}. 

Several techniques have recently been developed to handle this unprecedented amount of survey imaging data. These range from semi-automatic algorithms searching for arc and ring-shaped features \citep[e.g.][]{More2012, Gavazzi2014, Sonnenfeld2018}, to crowdsourcing science \citep[e.g.][]{Marshall2016, Sonnenfeld2020}. In this context, machine learning and deep learning methods appear to be a reliable and efficient mean to identify GGSLs (see, for example, the discussion in \citealt{Metcalf2019}), although they need to be trained on appropriate simulated datasets. In fact, the restricted number of confirmed strong-lensing examples in galaxy clusters prevents from training machine learning methods with real data. Moreover, the large redshift range over which GGSLs are searched for, their different morphologies, colours and magnitudes require realistic simulations to make deep learning-based methods effective in detecting real strong lenses. 

To this aim, significant efforts have been made over the last years to simulate the GGSL population as those observed by current and upcoming surveys. Mock images of strong-lensing events are obtained by coadding simulated lensed sources to foreground galaxies with different methods. For example, dark matter halos and galaxies can be extracted from semi-analytical catalogues (e.g. with the Millennium Observatory project, as done by \citealt{Metcalf2019}l, or by Leuzzi et al. in prep.), using mass density profiles \citep[e.g.][]{Collett2015, He2020, Lanusse2018}, or deep learning algorithms \citep{Lanusse2021}. Other studies opted for a hybrid approach which consists in modelling the mass density profile of photometrically selected galaxies \citep[e.g.][]{Petrillo2017, Petrillo2019, Li2020, Li2021, Gentile2022, Canameras2021, Akhazhanov2022}. Similarly, lensed sources can be simulated by modelling their surface brightness distributions \citep[e.g.][]{Petrillo2017, Petrillo2019, Li2020, Li2021, Gentile2022} or sampled from observations \citep[e.g.][]{Meneghetti2008, Meneghetti2010, Metcalf2019}, and then coadded to real or synthetic images through ray-tracing techniques (e.g. \texttt{GLAMER} \citealt{Metcalf2014, Petkova2014}, \texttt{GRAVLENS} \citealt{Keeton2001}).

In this work, we present a novel approach which exploits accurate cluster deflection fields to generate thousands of galaxy-galaxy strong lenses in galaxy clusters. The deflection angle maps are provided by high-precision cluster lens models, constructed by \cite{bergamini2019, Bergamini_m0416} and \cite{caminha2019} with the \texttt{LensTool} software \citep{Kneib1996, Jullo2007, Jullo2009}, by exploiting large numbers of spectroscopic multiple images. These models accurately describe both the cluster-scale mass component and the sub-halo mass distribution associated to the cluster galaxies, which together affect the morphology, brightness and frequency of galaxy-scale lensing events, for a given distribution of background sources. Thus, GGSLs can be simulated with a realistic description of the cluster galaxies acting as lenses in combination with the cluster-scale deflection field.

We test this methodology by injecting background source galaxies in multi-band images obtained with the Hubble Space Telescope (HST) as part of dedicated campaigns over the last decade, such as the Cluster Lensing And Supernova survey with Hubble \citep[CLASH,][]{postman2012}, Hubble Frontier Fields \citep[HFF,][]{Lotz2017} and Reionization Lensing Cluster Survey \citep[RELICS,][]{Coe2019}. 
This high-quality imaging dataset is completed with intensive spectroscopic programs, such as the CLASH-VLT \citep{Rosati2014} with VIMOS \citep[Visible MultiObject Spectrograph,][]{Lefevre2003} and MUSE observations \citep[Multi Unit Spectroscopic Explorer,][]{Bacon2012,bacon2014, Bacon2015} from the Very Large Telecscope (VLT), and GLASS \citep[Grism Lens-Amplified Survey from Space,][]{Treu2015, Schmidt2014}, which have offered a three-dimensional view of $\sim50$ clusters, providing spectra for several thousand galaxies. We exploit the combination of the imaging and spectroscopic datasets to construct our convolutional neural networks' knowledge base (KB).

This paper is structured as follows. In Sec.~\ref{sec:CNN}, we describe the two implemented convolutional neural networks. In Sec.~\ref{sec:method}, we illustrate the simulation methodology and dataset configuration. We detail the network performances in Sec.~\ref{sec:performance}, including a complete analysis of the network misclassifications as a function of the physical parameters. In Sec.~\ref{sec:run}, we test the generalisation capabilities acquired by the networks by processing a set of known GGSLs. Finally, we draw our conclusions in Sec.~\ref{sec:conclusions}.

Throughout the paper, we adopt a flat $\Lambda$CDM cosmological model with $\Omega_M$=0.3, $\Omega_\Lambda$= 0.7, and H$_0=70$\,km\,s$^{-1}$\,Mpc$^{-1}$. All astronomical images are oriented north to the top and east to the left. Unless otherwise specified, magnitudes are in the AB system.

\section{Convolutional neural network}\label{sec:CNN}
Among the deep learning methods, convolutional neural networks (CNNs, \citealt{LeCun:1989, LeCun1998}) have become well-established techniques to search for GGSLs in imaging surveys, owing to their ability to automatically extract information from raw data \citep[e.g.][]{Petrillo2017, Petrillo2019, Spiniello2018, Jacobs2019a, Jacobs2019b, Canameras2020, Huang2020, Li2020, Li2021, Gentile2022}. Here, we present the results achieved by two CNN architectures\footnote{We tested different network architectures, e.g. Residual~Net~X \citep{he2015, Xie:2016} and Inception Net \citep{szegedy:2014}. Due to their lower performances and higher computational cost, we limit the description of our results to deep models which achieved the best performances.}, both of which are inspired by the Visual Group Geometry network \cite{simonyan:2014}. The first network (similarly to the one described in \citealt{Angora2020}), hereafter named VGG, consists of a chain of convolution and pooling layers, whose ensemble of extracted feature maps are connected to the output through two dense layers. The second network consists of parallel VGGs, each of which processes a single HST band; hence, we name this architecture Single Channel VGG (SC-VGG). Since we use the F435W, F606W, F814W ACS bands in this work, the SC-VGG is composed of three parallel VGGs. Therefore, while the VGG performs a (linear) combination of filters in the first convolutional layer, the SC-VGG separates the informative contribution carried by the three bands. To obtain a single probability value, we average the probabilities for a GGSL event derived from each parallel VGG, which is used to measure the loss function and to update the training parameters. For both networks, we set the binary cross-entropy as loss function \citep{goodfellow:2016}, the Leaky version of the Rectified Linear Unit \citep[LeReLU,][]{Maas:2013} as the activation function for each layer and \texttt{Adadelta} \citep{Zeiler:2012} as optimiser. 

Furthermore, we include \textit{(i)} an \textit{early stopping} regularisation criterion \citep{Prechelt1997, Raskutti:2011} preventing overfitting; and \textit{(ii)} a gradual reduction of the learning rate on the plateau of the loss function \citep[as a function of iterations,][]{bengio:2012}. These techniques evaluate the network performance during the training phase, using a validation set previously extracted from the whole KB. At the same time, we opt for a stratified k-fold approach \citep{Kohavi1995, hastie2009} to handle the training-testing splitting, where a fraction of the training image cutouts is augmented through flipping and rotations, as described in \cite{Angora2020}.

Finally, to avoid memory loss, the networks have been trained with input data batches which include $32$ and $16$ patterns, respectively, for the VGG and the SC-VGG models. All networks have been implemented through \texttt{keras} \citep{keras2015}, with \texttt{tensorflow} \citep{tensorflow2015} as back-end system.

\section{Methodology}\label{sec:method}

\subsection{The simulation process}\label{ss:simulation}

\begin{figure*}[tb]
   \centering
   \includegraphics[width=\linewidth]{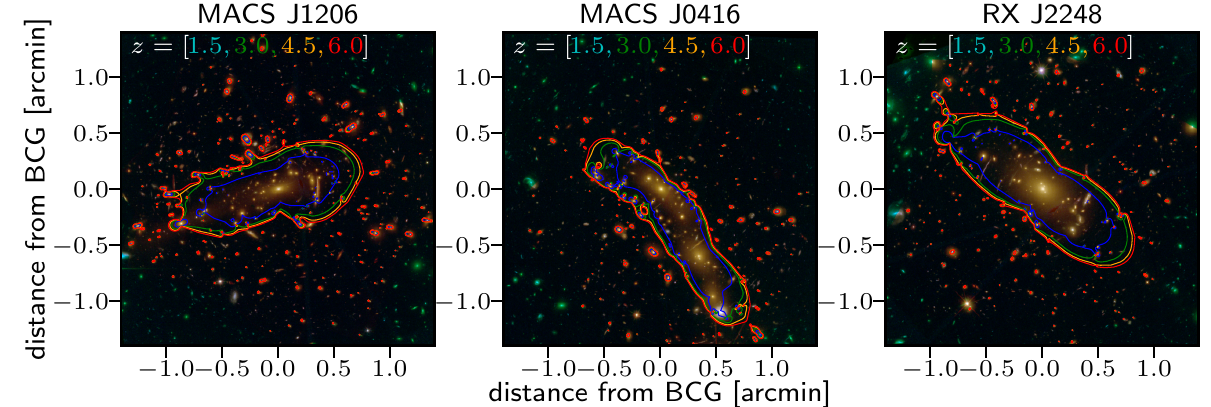}
   \caption{HST colour-composite images (combining the F435W, F606W, F814W filters) of three of the clusters in the sample (see Tab.~\ref{tab:GGSL:KB:clusters}). The tangential critical lines corresponding to $4$ different redshifts, $z=[1.5, 3.0, 4.5, 6.0]$, are shown in cyan, green, orange and red, respectively.} \label{fig:GGSL:clusters}
\end{figure*}

\begin{table*}[tb]\caption{Description of the cluster sample included in the GGSL simulation.}\label{tab:GGSL:KB:clusters}
\centering
\begin{tabular}{llccccccc}
\hline
Cluster & & $z_{cluster}$ & Survey & $M_{200c}^{(a)}[10^{14}M_{\odot}]$ & N$_\text{img}$ & N$_\text{CLM}$ (N$_\text{CLM}^\text{phot}$) & $\Delta_{rms}[\arcsec]$ & ref\\\noalign{\vskip 0.5mm}\hline
RX~J2129+0005 & R2129 & 0.234 & CLASH & 7.8$\pm$2.4 & 22 & 70 (34) & 0.20 & (1)\\
RX~J2248-4431$^{(b)}$ & R2248 & 0.346 & HFF & 19.8$\pm$6.0 & 55 & 222 (115) & 0.55 & (2)\\
MACS~J1931-2635 & M1931 & 0.352 & CLASH & 11.6$\pm$8.8 & 19 & 120 (59) & 0.38 & (1)\\
MACS~J0416-2403 & M0416 & 0.397 & HFF & 11.4$\pm$2.7 & 182 & 193 (49) & 0.40 & (3)\\
MACS~J1206-0847 & M1206 & 0.439 & CLASH & 15.1$\pm$3.2 & 82 & 258 (147) & 0.46 & (2)\\
MACS~J0329-0211 & M0329 & 0.450 & CLASH & 12.7$\pm$2.2 & 23 & 106 (49) & 0.24  & (1)\\
RX~J1347-1145 & R1347 & 0.451 & CLASH & 35.4$\pm$5.1 & 20 & 114 (70) & 0.36 & (1)\\
MACS~J2129-0741 & M2129 & 0.587 & CLASH & 1.84$\pm$0.01$^{(c)}$ & 38 & 138 (45) & 0.56 & (1)\\
\hline
\end{tabular}
\tablefoot{The first three columns list the: cluster names, short names and redshifts. The fourth column specifies the program from which the images are extracted. N$_\text{img}$ (Col.~5) is the number of multiple images used to constrain the model, N$_\text{CLM}$ (Col.~6) is the number of cluster members (CLMs) used to describe the sub-halo mass component (between brackets the number of CLMs photometrically selected), $\Delta_\text{rms}$ (Col.~7) is the root-mean-square separation between the observed and model-predicted multiple images positions. The reference lens model for each cluster is quoted in the last column. \\
$^{(a)}$ The cluster virial mass values have been measured through weak lensing by \cite{umetsu2018}.\\
$^{(b)}$ The cluster RX~J2248.7$-$4431 is also known as Abell~S1063.\\
$^{(c)}$ The weak lensing measurement is not available for M2129, we report here the mass within $200$\,kpc from \cite{caminha2019}.
}
\tablebib{
(1)~\cite{caminha2019}; (2)~\cite{bergamini2019}; (3)~\cite{Bergamini_m0416}.
}
\end{table*}

\begin{figure*}[tb]
   \centering
   \includegraphics[width=0.95\linewidth]{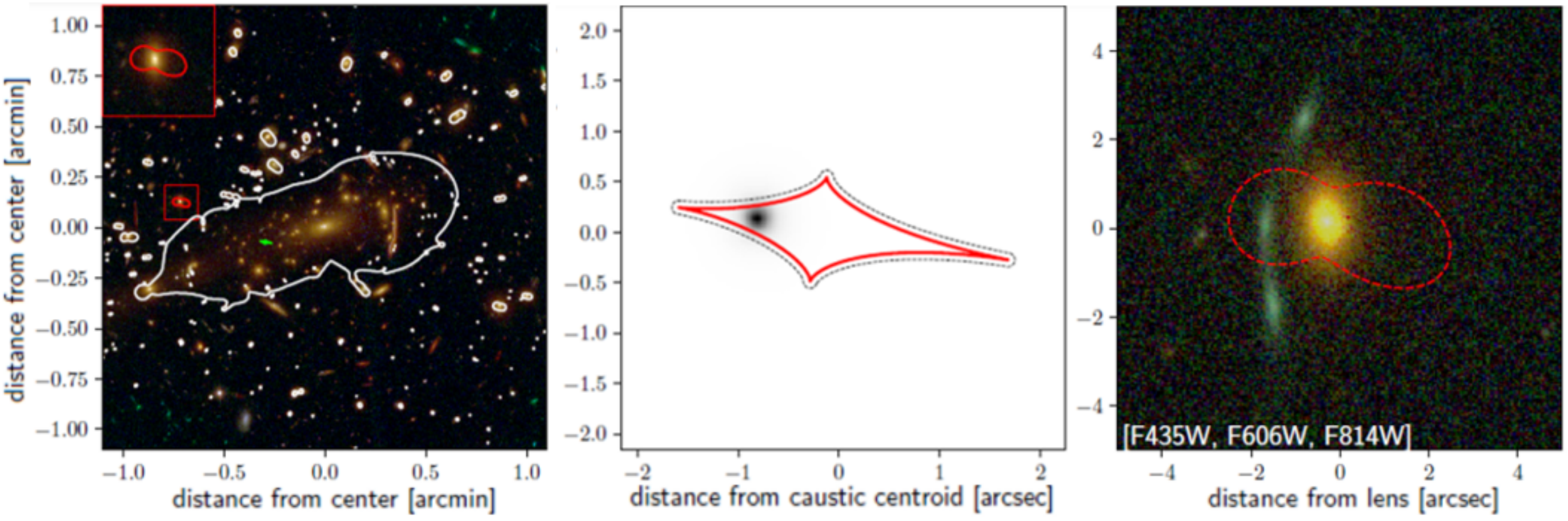}
   \caption{Example of a GGSL simulation. \textit{Left panel}: HST image of the cluster M1206 at $z=0.439$ ($\sim2\arcmin$ across), with the critical lines (in white) at $z=2.5$ from the lens model \citep{bergamini2019}. A specific secondary critical line is marked in red and zoomed-in in the upper left $\sim10\arcsec$ inset. The green spot indicates the position of the corresponding caustic on the source plane. \textit{Central panel}: source plane at $z=2.5$ showing the caustic (in red), corresponding to the selected critical line, including the buffer (black dotted line) delimiting the injecting region; the injected source has a \sersic profile (index $n=1.5$,  $R_{eff}=0.14\arcsec$), mag$_{F814W}=26.3$ and a start-forming galaxy SED. \textit{Right panel}: colour composite image of the resulting simulated GGSL system, together with the critical line (red dotted line, with a circularised $\theta_E=1.7\arcsec$); the cutout is $\sim10\arcsec$ across.
   }\label{fig:GGSL:example}
\end{figure*}

\begin{table*}[tb]\caption{List of \sersic parameters and their adopted value ranges for the injected sources.}\label{tab:GGSL:sersic_params}
\center
\begin{tabular}{lll}\hline
Parameter & Symbol & Extraction description \\\hline
Coordinate (source plane) & $y_s$ & Extracted within a buffer around the caustic (width $0.5\, r_e$)\\
Source magnitude & $m_{F814W}$ & Sampled from PDF, $p(i)$, COSMOS $+$ HST fields \\
Source redshift & $z_s$ & Sampled from PDF, $p(z|\Delta i)$, COSMOS \\
\multirow{2}{*}{Effective Radius} & \multirow{2}{*}{$R_e$} & $R_e = 2.54$\,kpc, $z\le1$\\
    &&$R_e(z) = B(1+z)^\beta$, $z>1$ \citep{Shibuya2015}\\
\sersic index & $n$ & Extracted within $(1.0,2.0)$\\
axis ratio & $q$ & Extracted within $(0.2,1.0)$\\
position angle & $\varphi$ & Extracted within $(0,\pi)$\\
\hline
\end{tabular}
\end{table*}

\begin{figure*}[tb]
   \centering
   \includegraphics[width=\linewidth]{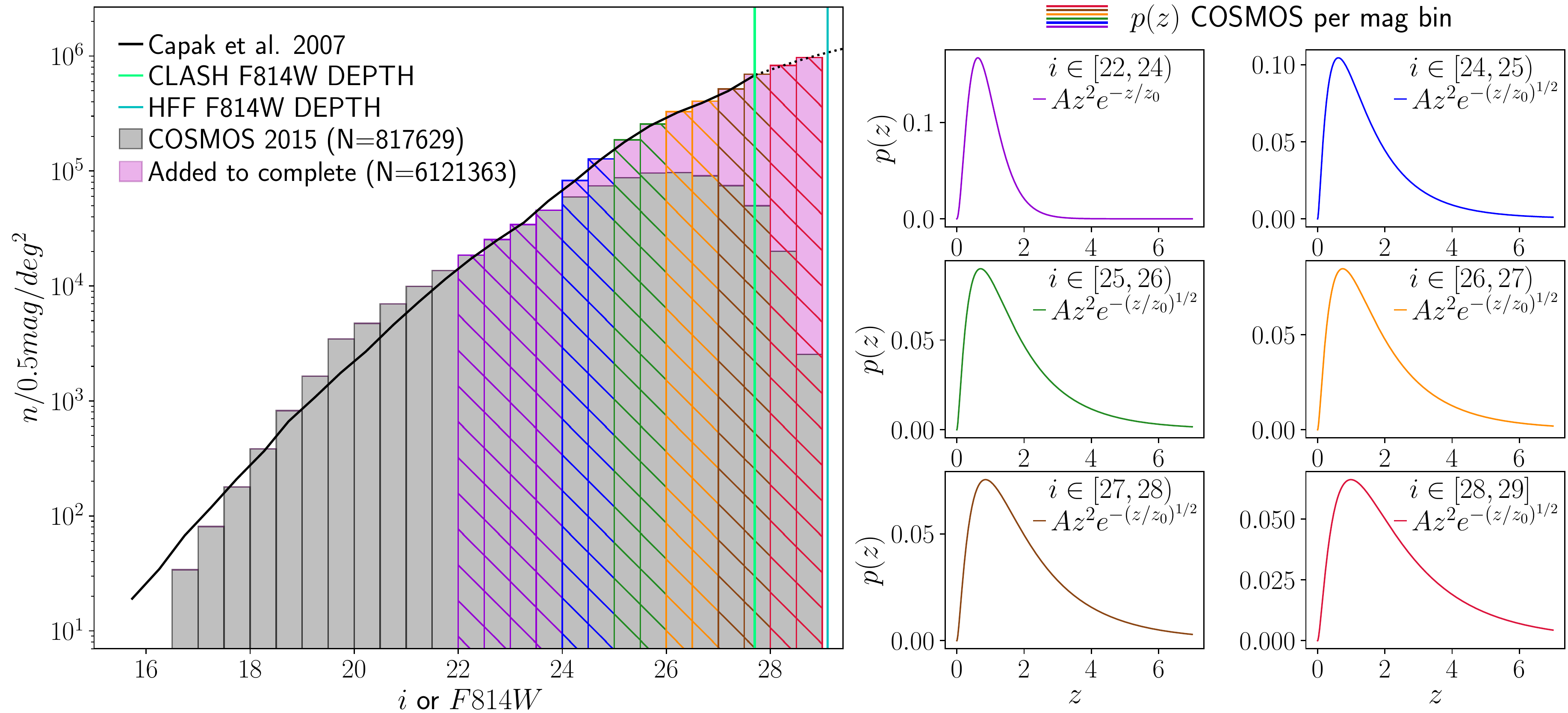}
   \caption{\textit{Left panel}: galaxy number counts estimated from the COSMOS $i$-band catalogue (grey bars), compared with \citep[][black line]{Capak2007}, together with the $5\sigma$ HFF and CLASH $F814W$ depth limit (cyan and green vertical lines). Galaxy counts added in the faint end to match the HST deep counts are coloured in magenta. \textit{Right panels}: redshift distributions for six magnitude bins, $p(z|\Delta m)$, coloured according to the magnitude bin from which they are extracted (left plot). }\label{fig:GGSL:numbercounts}
\end{figure*}

\begin{figure*}[tb]
   \centering
   \includegraphics[width=\linewidth]{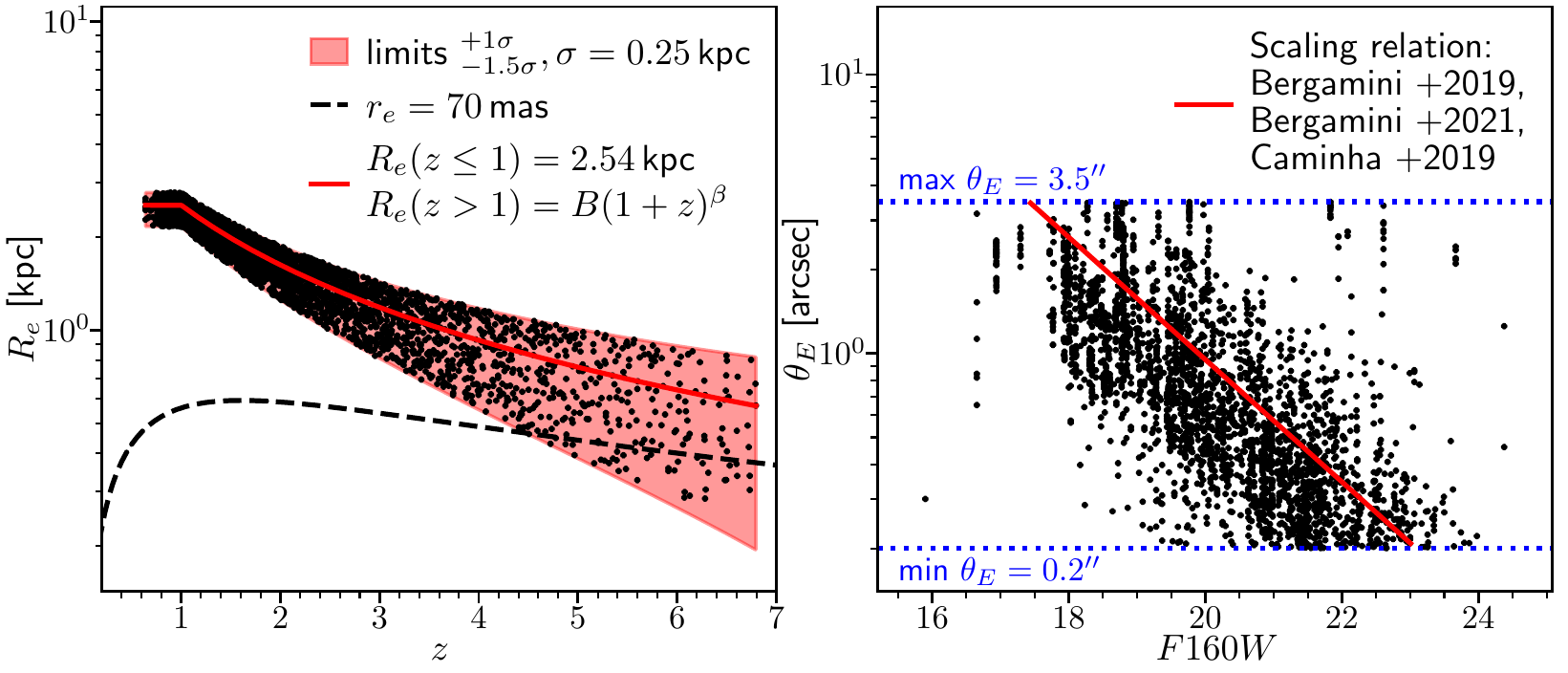}
   \caption{\textit{Left panel}: adopted relation for the redshift evolution of $R_e$, constant for $z\le1$ and taken from \cite{Shibuya2015} for $z>1$, together with the upper and lower limits within which $R_e$ is extracted (light red area). The black dashed line shows the $0.070\arcsec$ threshold, under which the source size is indistinguishable from the PSF, after the convolution. \textit{Right panel}: the resulting scaling relation, i.e. $\theta_E$ vs. $F160W$, compared to that from \cite{bergamini2019, Bergamini_m0416} and \cite{caminha2019}, in red.
   }\label{fig:GGSL:distributions}
\end{figure*}

\begin{table}[tb]\caption{Description of the cluster sample used in the non-GGSL selection.}\label{tab:non-GGSL:clusters}
\centering
\begin{tabular}{llcc}
\hline
Cluster & & $z_{cluster}$ & $N$ \\\hline
Abell~383 & A383 & 0.188 &  70              \\
Abell~209 & A209 & 0.209 &  75              \\
RX~J2129+0005 & R2129 & 0.234 & 51          \\
Abell~2744 & A2744 & 0.308 & 126            \\
MS~2137-2353 & MS2137 & 0.316 & 52          \\
RX~J2248-4431$^{(b)}$ & R2248 & 0.346 & 178 \\
MACS~J1931-2635 & M1931 & 0.352 & 28        \\
MACS~1115+0129 & M1115 & 0.352  & 96        \\
Abell~370 & A370 & 0.375 & 172        \\
MACS~J0416-2403 & M0416 & 0.397 & 120        \\
MACS~J1206-0847 & M1206 & 0.439 & 147       \\
MACS~J0329-0211 & M0329 & 0.450 & 66       \\
RX~J1347-1145 & R1347 & 0.451 & 44        \\
MACS~J1311-0310 & M1311 & 0.494 & 53       \\
MACS~J1149+2223 & M1149 & 0.542 & 130       \\
MACS~J2129-0741 & M2129 & 0.587 & 45       \\
\hline
\end{tabular}
\tablefoot{The cluster name, short name, and redshift are listed in the first $3$ columns. The fourth column shows the number of non-GGSLs identified through visual inspection.\\
$^{(a)}$ The cluster RX~J2248$-$4431 is also known as Abell~S1063.
}
\end{table}

To simulate the GGSL events, we exploit the deflection angle maps of eight galaxy clusters obtained from cluster lens models\footnote{The cluster lens models are publicly available at \url{https://www.fe.infn.it/astro/lensing/}.} provided by \cite{bergamini2019, Bergamini_m0416} and \cite{caminha2019}. The cluster sample is  described in Tab.~\ref{tab:GGSL:KB:clusters}, while three of the clusters are shown in Fig.~\ref{fig:GGSL:clusters}. 
The cluster total mass distribution of each cluster is modelled with a parametric description of the overall lensing potential, which includes a cluster-scale term composed of a dark matter halo and the smooth intra-cluster hot-gas mass from Chandra X-ray data, when available \citep{Bonamigo2017, Bonamigo2018}, and a clumpy component associated to the cluster member galaxies (CLMs). For the latter, the mass density profile of each sub-halo, containing both dark matter and baryons, is modeled with a circular, singular dual-pseudo isothermal profile \citep{Limousin2005, Eliasdottir2007} and further calibrated with the measured stellar velocity dispersions of a large samples of member galaxies  \citep{bergamini2021}. 
Such lens models are able to reproduce the observed positions of many multiple images (ranging from $\sim20$ to $\sim200$, see Table.~\ref{tab:GGSL:KB:clusters}) with a typical accuracy of $\lesssim 0.5\arcsec$. 

\texttt{LensTool} reconstructs the cluster potential by minimising the difference between the observed and model-predicted positions of the multiple images, given a set of model parameter values. The deflection angle maps, $\vec\alpha$, describe the relation between the source real position ($\vec\beta$) and its observed position ($\vec\theta$) via the lens equation: $\vec\beta = \vec\theta - \vec\alpha$. The simulation process is carried out with \texttt{PyLensLib} \citep{Meneghetti2021} and can be summarised as follows:
\begin{itemize}
    \item[-] From the deflection angle maps, we derive the convergence and the shear maps, i.e. the elements of the Jacobian matrix describing the image deformation, whose inverse matrix is the so-called \textit{magnification tensor}. Then, the critical curves are found where the magnification goes to infinity. Examples of tangential critical curves, corresponding to sources at four different redshifts, are shown in Fig.~\ref{fig:GGSL:clusters}, overlaid onto the HST field of view (FoV).
    \item[-] To avoid the primary critical lines associated to the cluster potential and very small-scale galaxies, we select the secondary critical lines whose equivalent (circularised) Einstein radius is  $0.2\arcsec < \theta_E < 3.5\arcsec$, which is consistent with the expected distribution of the equivalent Einstein radii associated to secondary critical lines in galaxy clusters (see, for example, Fig.~7 in \citealt{Meneghetti2022}). Moreover, we assign a selection probability proportional to $\theta_E$  (i.e. larger critical lines are more likely to be extracted). In this way, a mass-limited sample of lens galaxies is selected from the secondary critical lines, circumventing any photometric selection (see the left panel in Fig.~\ref{fig:GGSL:example}). 
    \item[-] The selected secondary critical line is mapped into the corresponding caustic on the source plane (see the central panel in Fig.~\ref{fig:GGSL:example}) using the lens equation.
    \item[-] The source is simulated by injecting a \sersic surface brightness profile \citep{sersic1963, sersic1968}, $I(\vec\beta)$,  within the caustic, including a buffer whose width is set equal to half of the source effective radius. Therefore, since the lens mapping conserves the surface brightness, i.e. $I(\vec\theta) = I(\vec\beta)$, the observed surface brightness is computed as $I(\vec\theta = \vec\beta + \vec\alpha)$. The resulting GGSL is finally generated by convolving the simulated multiple image system with the HST point spread function (PSF) for each band and then coadded to the HST ACS image in a given filter (right panel in Fig.~\ref{fig:GGSL:example}). The used PSFs are estimated with \texttt{morphofit} (\citealt{Tortorelli_morphofit}, see also \citealt{Tortorelli2018, Tortorelli2023}).    
\end{itemize}

\noindent
In this work, we adopt a source spectral energy distribution (SED) of a star-forming galaxy template from \cite{Kinney1996}. 
The list of \sersic parameters and their adopted value range are shown in Tab.~\ref{tab:GGSL:sersic_params}. The \sersic index is extracted from a uniform distribution between $n=1.0$ and $2.0$, typical of late-type galaxy star-burst profiles. The axis ratio and the position angle values are randomly extracted from uniform distributions in $(0.2, 1.0)$ and $(0, \pi)$, respectively. To closely reproduce the HST observations, we do not use a uniform sampling for the other parameters. Specifically, for the source magnitudes and redshifts, we estimate the number counts in the $i$-band (i.e. the number of galaxies per square degree per magnitude bin) from the COSMOS~2015 catalogue \citep{Scoville2007, Laigle2016}, complemented with HST Deep Field North and South observations \citep{Williams1996, Metcalfe2001} in $F814W$ (taken from \citealt{Capak2007}), which extends the galaxy counts to the faint end, down to $F814W=29$\,mag, (see the left panel of Fig.~\ref{fig:GGSL:numbercounts}). In each of the six magnitude bins (with $i$ limits $=\{22, 24, 25, 26, 27, 28, 29\}$\,mag), we use the COSMOS photometric redshift catalogue to estimate a redshift probability density function (PDF), i.e. $p(z\, |\Delta i)$, by fitting it with a simple function of the form: $p(z\, |\Delta i) = A z^2 e^{-z/z_0}$ for $i \in [22,24)$ and $p(z\, |\Delta i) = A z^2 e^{-(z/z_0)^{1/2}}$ for the other magnitude bins \citep[see, e.g. ][]{Lombardi1999, Lombardi2005}. The redshift limit of the COSMOS catalogue is $z\sim7$, appropriate for our studies in which the reddest band is $F814W$. The six modelled PDFs are shown in the right panels of Fig.~\ref{fig:GGSL:numbercounts}.  
For a given total number of galaxies to inject in the cluster field, appropriate for the depth of the HST observations, we then use these PDFs to assign a source magnitude and redshift to each background galaxy. We also impose a minimum value for the source redshift $z_\text{src} = z_{\text{cls}}+0.4$, as suggested by \cite{Meneghetti2022}, who measured the lensing cross-section for the galaxy clusters considered in this work, finding that it becomes significantly larger than zero for $z_\text{src} \gtrsim z_\text{cls}+0.4$.

Finally, to assign an effective radius value to the background galaxies, we exploit an empirical relation describing the redshift evolution of galaxy physical sizes proposed by \cite{Shibuya2015}, approximated with a function of the form: $R_e = B(1+z)^\beta$ (fitted by combining galaxy radii estimated in the UV and optical bands). However, since a comparison of these values with the effective radii measured by \cite{Tortorelli2018} for low-$z$ galaxies shows a significant overestimate, we limit the application of this relation only at $z>1$, adopting a constant value at $z\le 1$ (see the fourth row in Tab.~\ref{tab:GGSL:sersic_params} and the left panel in Fig.~\ref{fig:GGSL:distributions}). As suggested by the \cite{Shibuya2015} analysis, we assume a scatter $\sigma=0.25\,$kpc over the entire redshift range, and randomly extract a value of $R_e$ at a given redshift $z$ within the $[-1.5\sigma, +1 \sigma]$ range (see the left panel in Fig.~\ref{fig:GGSL:distributions}). 
The chosen asymmetrical range allows us to sample $R_e$ values down to $R_e\lesssim 0.5$\,kpc at $z\gtrsim4$ (as shown in the left panel of Fig.~\ref{fig:GGSL:distributions}).

In the effort to verify that our simulated galaxy-scale lenses statistically reproduce the observations, 
we compare the $\theta_E - m_{F160W}$ relation obtained for our mock GGSL sample with the cluster member velocity dispersion scaling relation, measured in \cite{bergamini2019, Bergamini_m0416} and used to build the lens models, i.e. $\sigma^\text{CLM}_i = \sigma^\text{ref} (L_i/L^\text{ref})^\alpha$ (see also \citealt{Brainerd1996} and \citealt{Jullo2007}). To this aim, we compute the expected Einstein radius as a function of the lens galaxy $F160W$ magnitude by assuming a singular isothermal sphere for the lens galaxy mass density profile \citep{Schneider2006}:
\begin{equation}
\nonumber
    \theta_{E, i} = 4\pi \Bigg(\frac{\sigma_v^\text{ref}}{c}\Bigg)^2 \Bigg(\frac{D_{LS}}{D_S}\Bigg)
     10^{\displaystyle0.8\,\alpha\,\Big(m^\text{ref}_{F160W} - m^\text{CLM}_i\Big)},
\end{equation}
where $m^\text{ref}_{F160W}$ is the $F160W$ reference magnitude, corresponding to the brightest cluster galaxy (BCG); $\sigma_v^\text{ref}$ is a free parameter of the lens model (the normalization of the $\sigma-m$ scaling relation); $D_S$ is the angular diameter distance to the source, and $D_{LS}$ is that between the lens and the source. The value of the slope of the scaling relation, $\alpha$, is the one used in the lens models (directly inferred from the stellar velocity dispersion measurements \citep{bergamini2019,Bergamini_m0416}. In Fig.~\ref{fig:GGSL:distributions}, we show the latter relation as a red line and remark that it closely follows the distribution of effective Einstein radius values inferred from the secondary critical lines.

\subsection{Building the knowledge base}\label{ss:KB}
The described methodology can simulate an arbitrary number of realistic GGSLs embedded in the complex environment of galaxy clusters, as observed with the HST. To build a KB containing a large variety of GGSLs, we generate twice as many mock GGSLs as non-GGSLs (NGGSLs, i.e. the negative class for the classification problem). To this aim, we exploit the spectroscopic information obtained by combining the CLASH-VLT VIMOS programme with the MUSE archival observations and extract $10\arcsec$ cutouts centered on the CLM positions, with a rest-frame velocity separation of $|v|\le 5000$\,km\,s$^{-1}$, belonging to $16$ clusters (see Tab.~\ref{tab:non-GGSL:clusters}). Since some of these cutouts may contain  strong lensing features, we carried out a visual inspection process by lensing experts in our group to build a fiducial sample of non-GGSLs. To help this classification process, each RGB image cutout was inspected together with the $F435W$, $F606W$, $F814W$ bands, with knowledge of any nearby spectroscopic source with $z_s\ge z_\text{cluster}+0.1$. A score of $+1$, $+0.5$, or $-1$ was assigned to each CLM in case of a reliable GGSL, a less likely GGSL, or a non-GGSL, respectively. The cutouts containing bright stars, nearby bright galaxies or with an incomplete multi-band coverage near the edge of the FoV were also excluded. In this way, using the average scores, a visual inspection of $16$ galaxy clusters led to the classification of a pure sample $1453$ non-GGSL. At the same time, $320$ were classified as candidate GGSLs, and $282$ cutouts were excluded. 

Therefore, the resulting KB comprises $1453$ observed non-GGSLs and $3000$ simulated GGSLs. This initial mismatch, motivated by the need for a sufficient diversity of mock GGSLs, is later compensated during the extraction of the validation subset and the augmentation process, which leads to a training set of $\sim3800$ images for each class. The KB dataset is then built by extracting $128\times128$ pixel cutouts ($3.84\arcsec$ by side)\footnote{We also studied the network behaviour using $256\times256$ pixel cutouts ($7.68\arcsec$ by side) and find out that the performances are worse ($8\%-15\%$ in terms of accuracy), in addition to a four times higher training computing time. These tests suggest that the $\sim4\arcsec$ cutouts offer the best strategy.}. By studying the distribution of the distances of the multiple images with respect to lens centers, we find that all the cutouts contain at least one lensed image.

\begin{figure*}[h]
   \centering
   \includegraphics[width=0.9\linewidth]{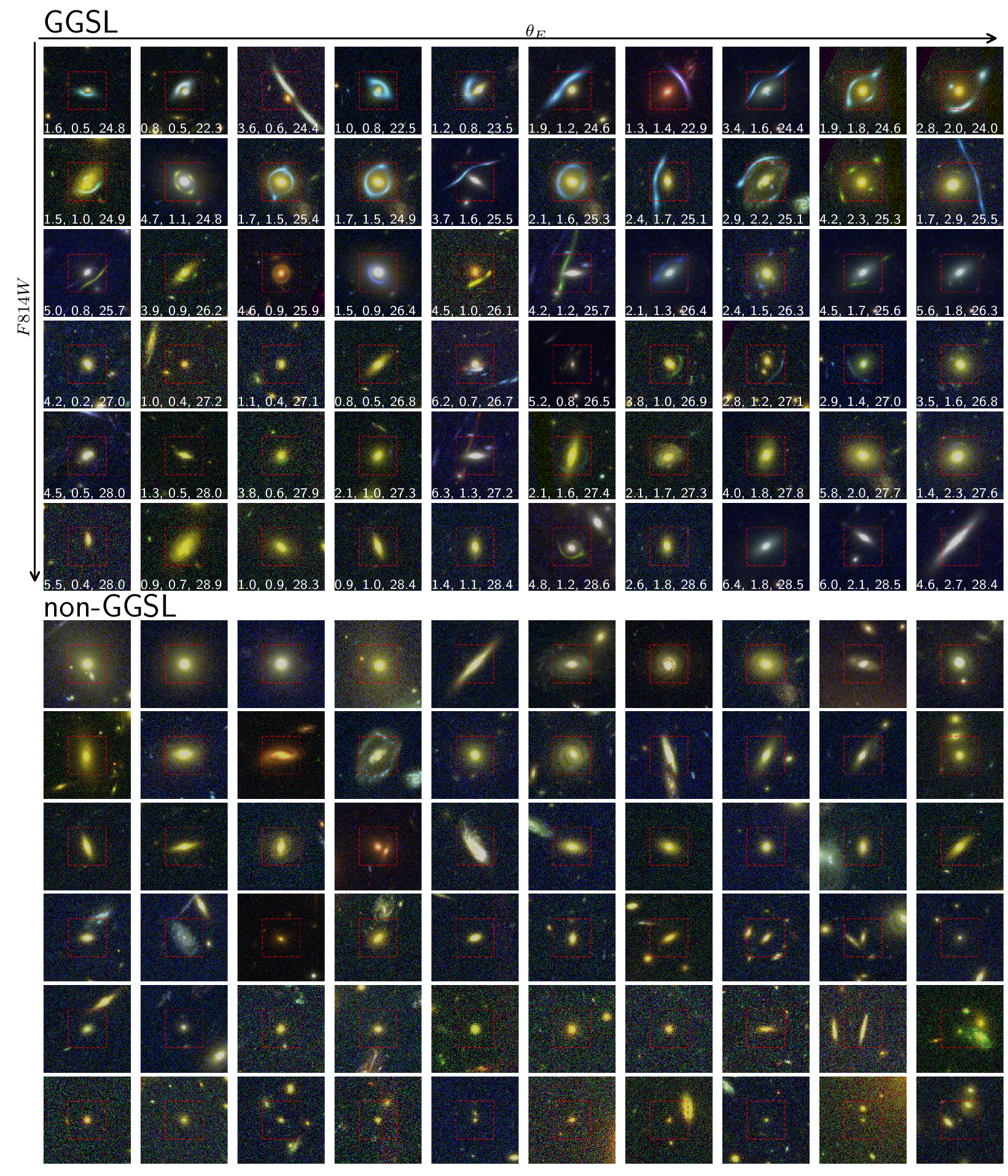}
   \caption{Examples of RGB cutouts of GGSLs and non-GGSL obtained by combining the $F435W$, $F606W$, $F814W$ bands. GGSL cutouts are sorted in order of increasing $\theta_E$ (along columns) and $F814W$ magnitude (along rows) values. The images have been stretched to emphasise faint features by clipping values within $\pm3\sigma$ and normalising them. Cutouts are $\sim9\arcsec$ across; red squares indicate the $4\times4\arcsec$ areas which are processed by the networks. The labels at the bottom of each image indicate the values of $z_s$, $\theta_E$ and $F814W$ magnitude.}\label{fig:GGSL:cutouts}
\end{figure*}

A sample of simulated GGSLs and cutouts classified as non-GGSLs is shown in Fig.~\ref{fig:GGSL:cutouts}, where the input images are indicated as red squares. GGSLs are sorted in order of increasing $\theta_E$ (across columns) and source intrinsic $F814W$ magnitude (across rows). Besides the typical arc-like and ring-like features, several GGSL mock images do not reveal any apparent strong lensing feature. This may occur (\textit{i}) when the injected source is too faint, so that the lens galaxy outshines the GGSL signal ($30\%$ of sources have $F814W>28$\,mag); (\textit{ii}) for small-scale lenses (small $\theta_E$), where the lens galaxy halo hides multiple images ($32\%$ of lenses have $\theta_E<0.5\arcsec$); (\textit{iii}) or a combination of the previous two cases ($10\%$ of GGSLs have both $F814W>28$\,mag and $\theta_E<0.5\arcsec$). Although these cutouts represent the most challenging cases for the classifier, they act as adversarial examples \citep{Szegedy2013}, preventing network overfitting and allowing the network to gain a high degree of generalisation \citep{Goodfellow2014_adversarial, Zhao2020, Kong2020}. We also performed some experiments to verify this by removing faint sources and small-scale lenses from the training phase. Even though the network achieves nearly perfect results, it appears unable to identify \textit{real} strong lensing events, lacking enough generalisation capabilities. Finally, we note that the same CLM cutout can be used in the training set as a mock GGSL or a non-GGLS (i.e. no background source is injected); however, we expect this to have a negligible impact on the CNN performance.

\section{Network performances}\label{sec:performance}

\subsection{Statistical metrics}\label{ss:metrics}
In order to assess the network classification performance, we use a set of metrics that are computed from the binary confusion matrix \citep[][see middle panel of Fig.~\ref{fig:vggs_comp}]{stehman:1997}, namely the average efficiency (AE), purity (\textit{pur}), completeness (\textit{compl}), and the \textit{F1}-score, which is the harmonic mean between purity and completeness. The accuracy represents a global average score which includes both classes, while the other three estimators are measured for each class. In this work, we refer to the GGSLs as the ``positive'' class. Therefore, the four elements of a binary confusion matrix assume the following meaning: true positives (TPs) are GGSLs correctly classified, false positives (FPs) are non-GGSLs incorrectly flagged as GGSLs, false negatives (FNs) are GGSLs wrongly predicted as non-GGSLs, true negatives (TNs) are non-GGSLs which are correctly classified. These estimators, as well as the confusion matrix, are computed by assuming a given threshold on the probability assigned by the CNN to each object. Unless otherwise specified, such a probability threshold is set to $0.50$. The performance assessment is completed by including the so-called receiver operating characteristic curve \citep[ROC,][]{hanley:1982}, which represents the trade-off between the true positive rate (TPR, i.e. the completeness rate) and the false positive rate (FPR, i.e. the contamination rate) as a function of the probability threshold (see the middle panel in Fig.~\ref{fig:vggs_comp}). The area under this curve (AUC) can be used as an additional estimator of the network classification capabilities. Finally, since the training-test split has been implemented with a $k$-fold approach, we can also analyse the metrics fluctuations over the $10$ folds (see \citealt{Angora2020}) and represent them graphically as in the bottom panel of Fig.~\ref{fig:vggs_comp}. For each metric, the box delimits the $25$th and $75$th percentiles, i.e. the first and third quartile ($Q_1$ and $Q_3$); their difference, the so-called interquartile range, $IQR = Q_3-Q_1$; the error bars (ranging from $Q_1-1.5\cdot IQR$ to $Q_3+1.5\cdot IQR$) correspond to the $90.3\%$ of data (i.e. within $\pm2.698\sigma$ values); the horizontal line indicates the median value.

\subsection{Performances}\label{ss:performances}

\begin{figure}[ht]
   \centering
   \includegraphics[width=0.95\linewidth]{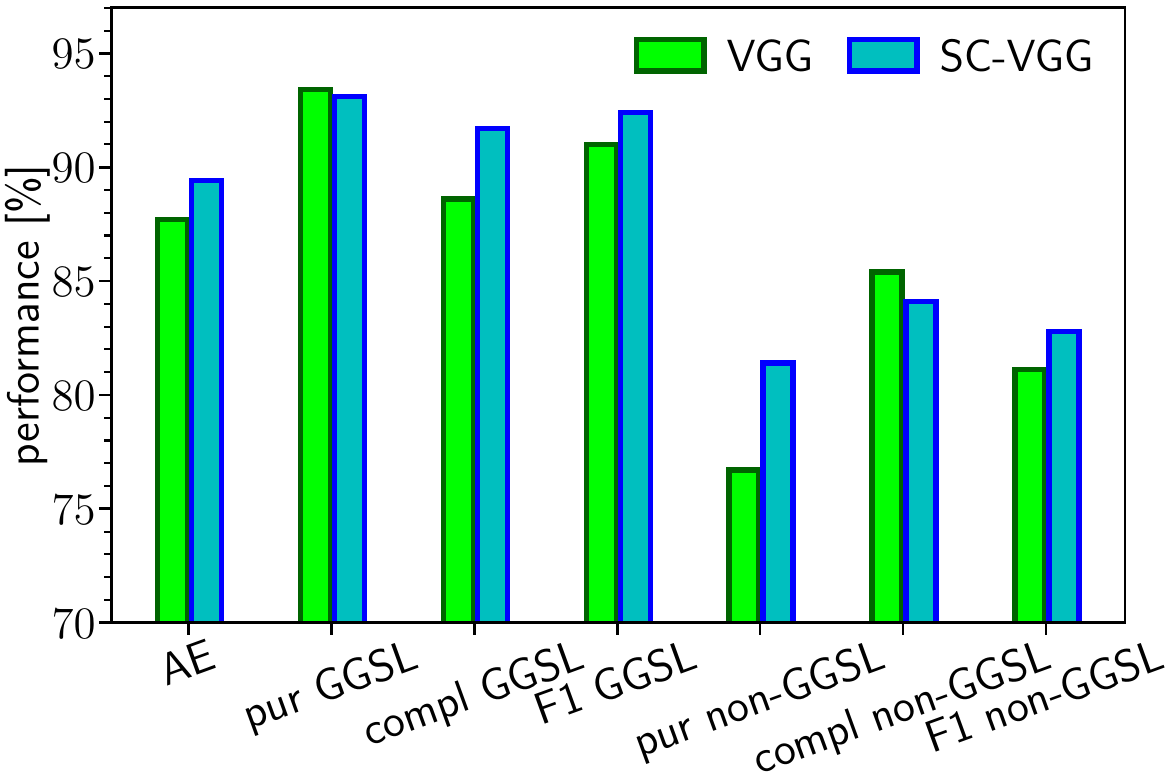}
   \includegraphics[width=0.95\linewidth]{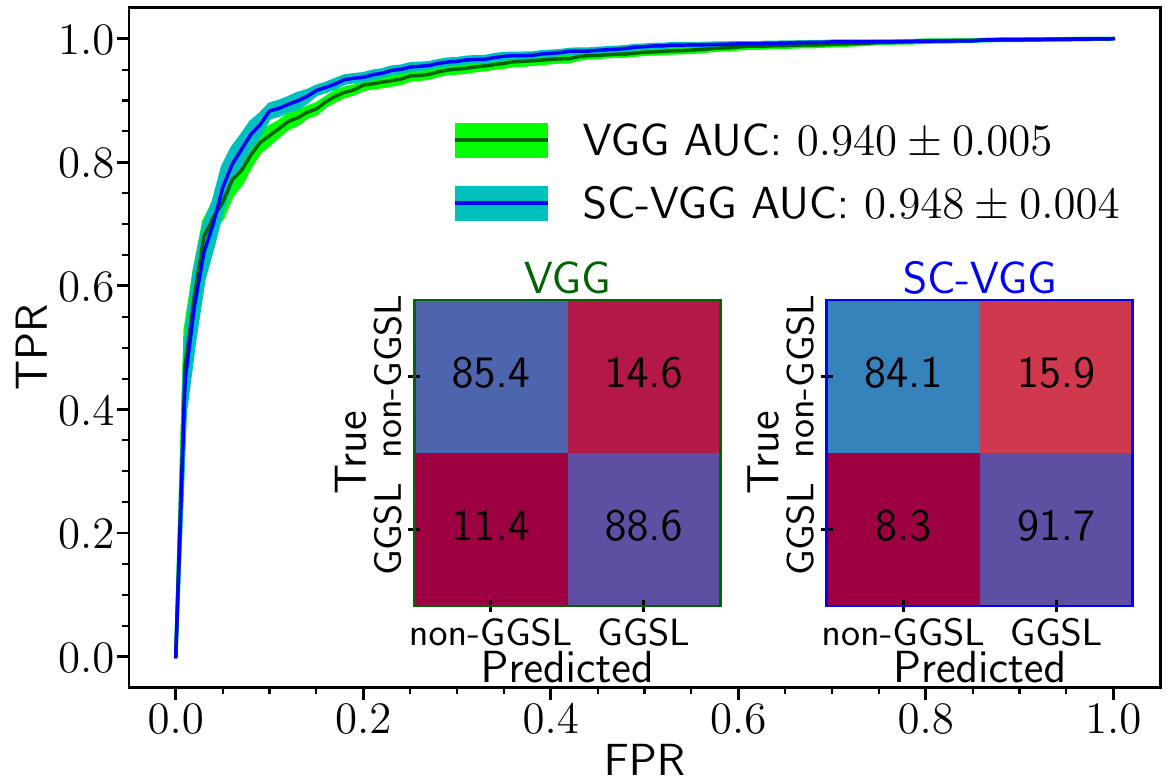}
   \includegraphics[width=0.95\linewidth]{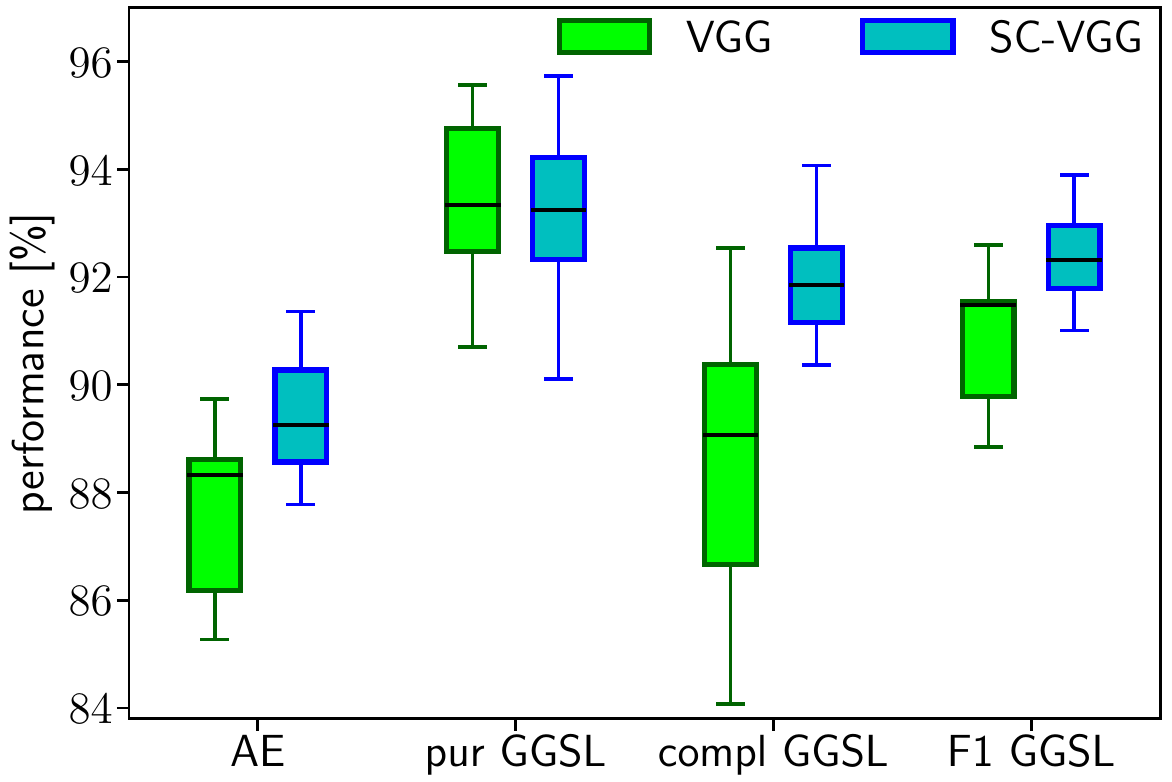}
   \caption{Comparison between the performance of the two networks under study (in all panels the VGG and SC-VGG results are shown in green and in cyan, respectively). \textit{Top panel}: statistical performance estimators for the GGSL and non-GGSL classes. \textit{Middle panel}: ROC curves for the GGSL classification, i.e. true positive rate vs false positive rate (the lines and the coloured areas represent the mean and the $1\sigma$ level, respectively), the AUC (area under the curve) values are quoted in the legend; the normalised confusion matrices are also shown. \textit{Bottom panel}: box plots for the GGSL metrics and average efficiency (AE) for both classes (see Sec.~\ref{ss:metrics} for details).}\label{fig:vggs_comp}
\end{figure}

A summary of the performances is shown in Tab.~\ref{tab:vggs_comp} and in the top panel of Fig.~\ref{fig:vggs_comp} in terms of the statistical estimators (purity, completeness, F1-score and average efficiency) for both classes. Globally, CNNs correctly classified at least  $87\%$ of the sources. Concerning the GGSL identification, both CNNs appear more pure than complete, with \textit{pur}$-$\textit{compl} differences ranging from $1.4\%$ to $4.8\%$. As for the non-GGSLs, the networks reveal an opposite behaviour and a wider trade-off (\textit{pur}$-$\textit{compl} between $-8.7\%$ and $-2.7\%$). Such a dichotomy results from an unclear distinction between the two classes for a fraction of sources, which in our case is represented by faint sources and small-scale lenses (i.e, the adversarial examples). As mentioned above, these images prevent model overfitting when included in the training set; however, we measure the statistical estimators with and without these adversarial images in the test sets to quantify their effect on estimating the model performances. These re-estimated metrics (i.e. without the adversarial cutouts) are marked with an asterisk in Tab.~\ref{tab:vggs_comp}. Clearly, the non-GGSL completeness is not affected by this modification, while the non-GGSL purity increases by $15.6\%$ and $11.9\%$, reaching $92.3\%$ and $93.3\%$, respectively, for the VGG and SC-VGG models. Correspondingly, the F1-score increases to $\sim88\%$, with an improvement of $7.6\%$ for the VGG and $5.7\%$ for the SC-VGG. Regarding the GGSL set, we find a more balanced trade-off between purity and completeness: the drop in purity ($\sim6\%$) is balanced by a gain in completeness ($\sim3\%$); while the F1-score drop remains within $\lesssim2\%$. 

A further analysis of the CNN performances is illustrated in Fig.~\ref{fig:vggs_comp}. In the middle panel, the ROC curves and the corresponding AUC values are similar (within $1\%$), whereas the bottom panel better details the network classification capabilities, quantifying the performance fluctuations (values listed in Tab.~\ref{tab:quantiles}). Both networks have similar GGSL purity (median, $\sim93\%$, first and third quartile, $Q_1\sim92\%$ and $Q_3\sim94\%$, and inter-quartile range $IQR = 2.1\%$). More significant variations occur for the other GGSL metrics: the SC-VGG performances show an overall significant improvement in terms of completeness (median: $2.8\%$, $Q_1$ and $Q_3$: $4.5\%$, $2.1\%$), which in turn is reflected into an F1-score gain (from $0.8\%$ to $3.8\%$). Concerning the non-GGSL metrics (only listed in Tab.~\ref{tab:quantiles}), SC-VGG achieves larger purity values ($4.7\%$), while VGG shows better completeness ($1.0\%$). SC-VGG achieves the best non-GGSL F1 score, with an average improvement of $1.6\%$. 

Based on this analysis, SC-VGG shows the best purity-completeness trade-off for both GGSL and non-GGSL ($92.4\%$ -- $82.8\%$ vs. $91.0\%$ -- $81.1\%$), it appears more robust when dealing with adversarial examples and is less subject to metric fluctuations ($\langle IQR \rangle_{SC-VGG}=2.1\%$ vs $\langle IQR \rangle_{VGG}=3.4\%$), particularly for the GGSL completeness.

Furthermore, we also perform an experiment using single-band cutouts (the $F435W$, $F606W$ and $F814W$ independently), i.e. removing the multi-band information. The results are outlined in Tab.~\ref{tab:singleband}, to be compared with the performances of the VGG and SC-VGG models. Although single-band performances reproduce the VGG and SC-VGG purity-completeness trends, the use of single-band data implies a loss of performance based on all metrics: an average reduction of $1.8\%$, $1.3\%$ and $3.5\%$, respectively for the AE, GGSL and non-GGSL F1-scores. However, this moderate performance loss suggests that GGSLs can also be classified using single bands when multi-band imagining is not available (see also \citealt{Petrillo2017}, \citealt{Petrillo2019} \citealt{Li2021}, who use the Kilo-Degree Survey data by \citealt{KiDS2015}). When using single band information, our tests show a better performance with the blue filter, owing to the larger contrast between the lens galaxy (red) and the strong-lensing features (generally blue).  

We also tested the network performance using a KB built with cutouts twice as large ($256$ pixels $\simeq7.7\arcsec$ side). These experiments show a $10\%$ drop in the performance metrics. Considering also the significant extra burden of computing resources, we did not pursue this strategy further. 

Finally, we compare the predictions made by our neural networks with the outcome of the visual inspection by gravitational lensing experts. As pointed out, since we aim to produce a highly pure non-GGSL sample, the resulting set of GGSL candidates is strongly contaminated, as it includes objects with uncertain visual classification (conservatively excluded from the non-GGSL set). Thus, a human-machine comparison is more appropriate for identifying non-GGSLs rather than GGSLs. Indeed, by considering the $\sim1800$ visually inspected sources, we measure a high fraction of non-GGSL predicted in common ($\sim95\%$), whereas the percentage of common GGSLs is just $\sim35\%$. However, by increasing the CNN probability threshold from $0.50$ to $0.75$, we find that all the $105$ candidates are classified as GGSLs by both neural networks and astronomers, underscoring the high effectiveness of the CNN developed in this work. 

\subsection{False positives and false negatives}\label{ss:FPFN}
\begin{figure*}[tb]
   \centering
   \includegraphics[width=0.95\linewidth]{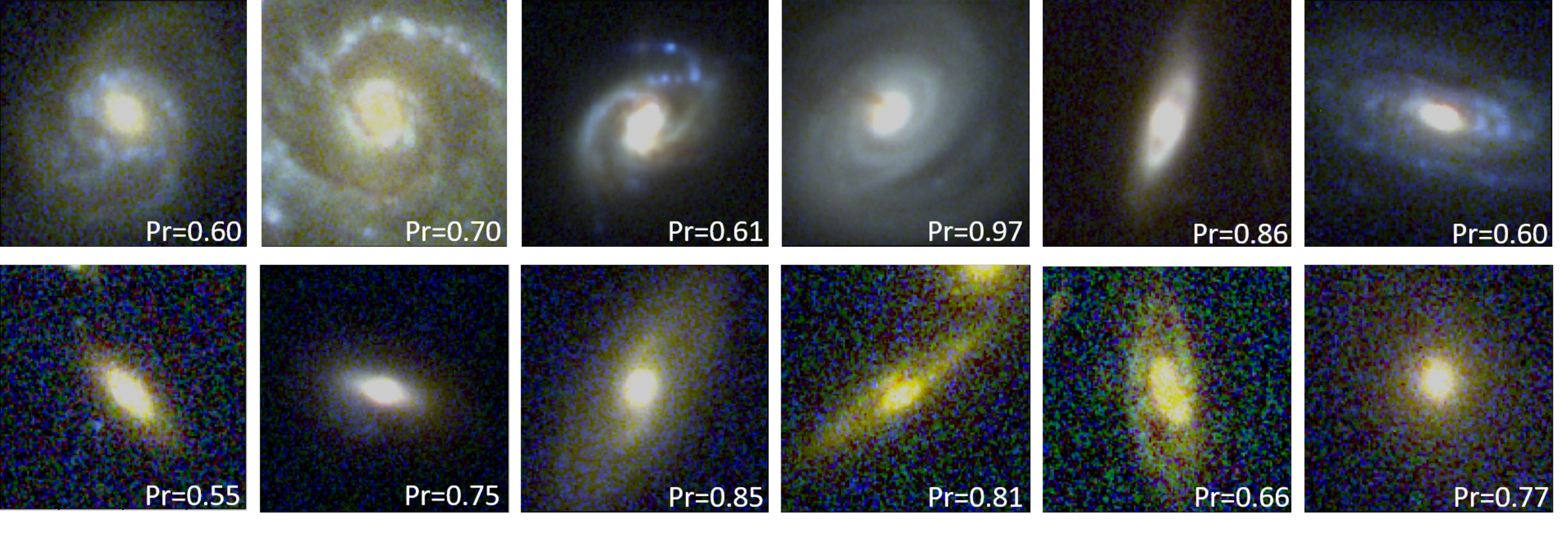}
   \caption{Selection of false positives common to both the VGG and SC-VGG models. The probability of belonging to the GGSL class is shown in each thumbnail (referred to the SC-VGG model). Cutouts are $\sim4\arcsec$ across.}\label{fig:FP:imgs}
\end{figure*}

\begin{figure*}[tb]
   \centering
   \includegraphics[width=0.48\linewidth]{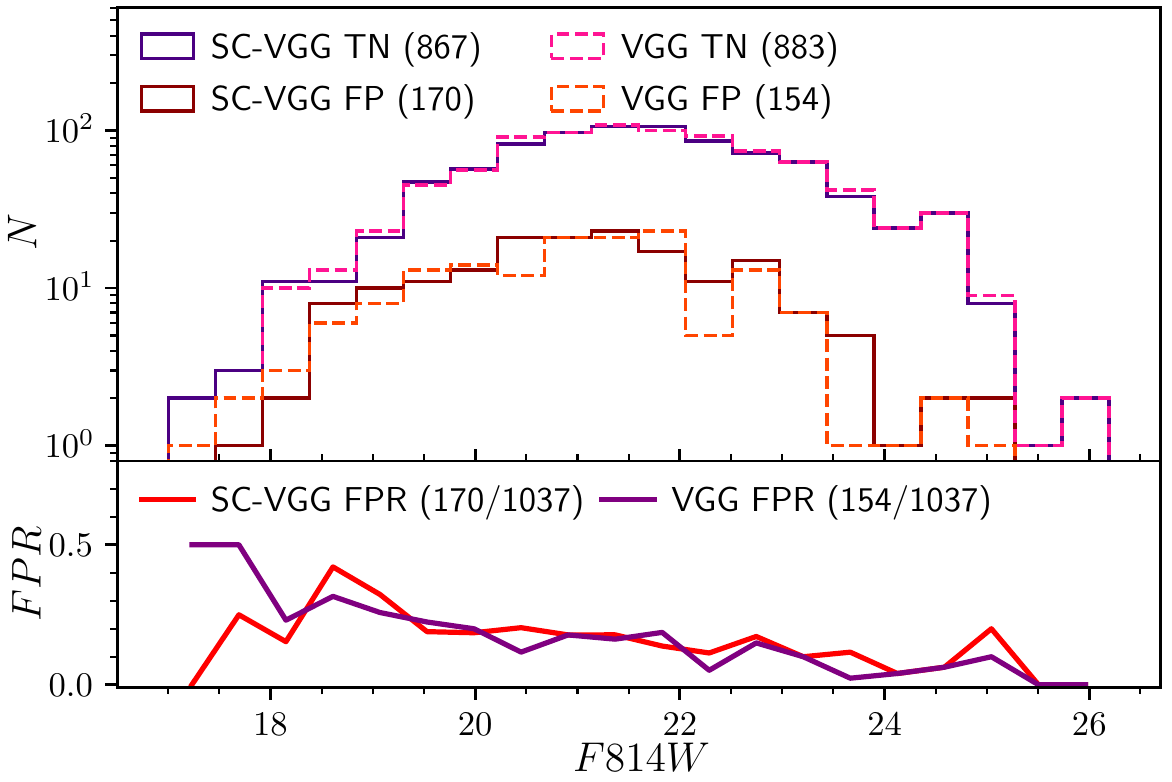}
   \includegraphics[width=0.48\linewidth]{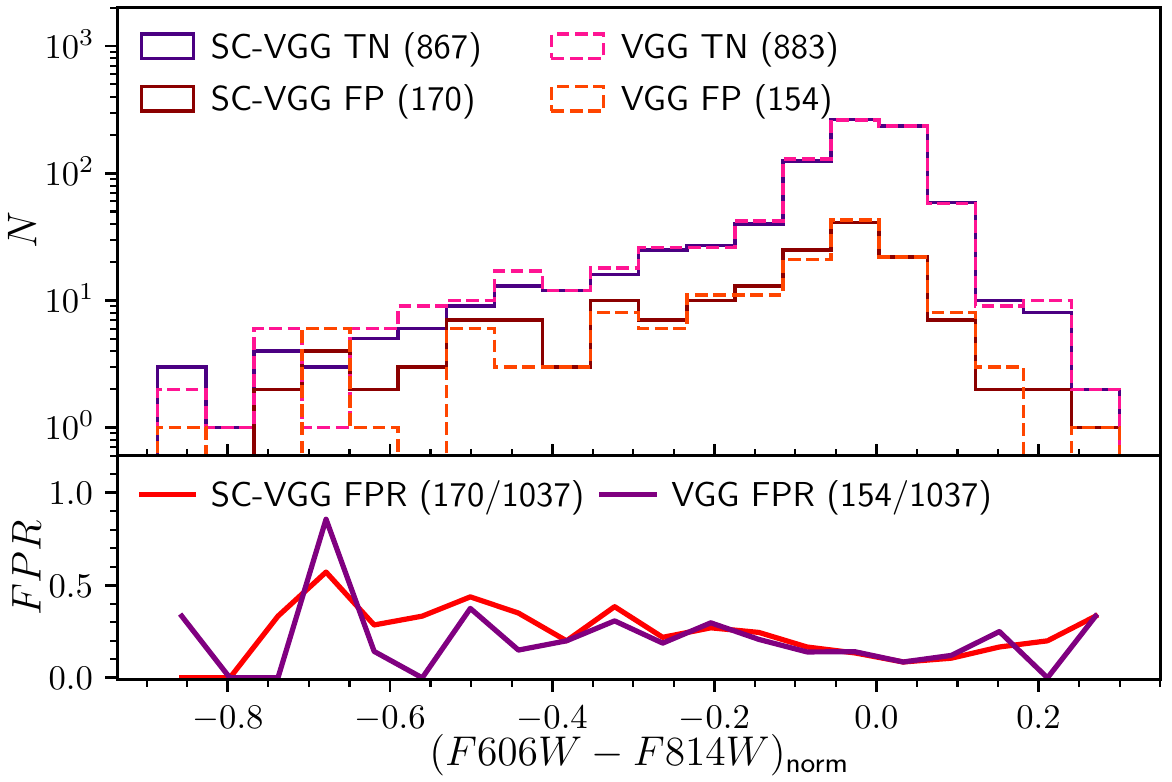}
   \caption{True negative (TN) and false positive (FP) analysis related to the VGG and SC-VGG performances, as a function of the lens galaxy lens photometry: $F814W$ magnitude (left panel), $(F606W-F814W)_{\text{norm}}$ normalised colour (right panel). In both panels, the TN rates are plotted with purple and magenta lines, while the FP rates with red and orange lines (respectively for the SC-VGG and VGG), in both cases, solid for SC-VGG, dashed for VGG. The FP ratio is plotted at the bottom of each panel (as a purple line for VGG, red for SC-VGG). In both panels, only sources with available and reliable magnitudes are plotted.}\label{fig:alx_vgg:FP}
\end{figure*}

In this section, we specifically analyse the properties of the false positives and false negatives produced by the CNN, which are characterised based on the galaxy magnitude and colour of FPs, and the GGSL system properties of FNs (source redshift and intrinsic magnitude, together with the lens Einstein radius). 

Concerning the non-GGSL mistakenly classified as strong-lenses, a selection of false positives common to both VGG and SC-VGG models is displayed in Fig.~\ref{fig:FP:imgs}. In Fig.~\ref{fig:alx_vgg:FP}, we show the TN, FP and the false positive ratio ($FPR=\frac{FP}{TN+FP}$) as a function of the cluster member photometry: $F814W$ magnitude (left panel) and the normalised colour (right panel), whose values are summarised in Tab.~\ref{tab:fp_dep}. We use the galaxy red-sequence dependence on the redshift to compensate for the K-correction of CLMs, thus obtaining a normalised colour. We use the \cite{girardi2015} relation, $(F814W - F606W)_{\text{norm}} = (F814W - F606W)_{\text{obs}} - \text{CM}(F814W)$, that is the difference between the observed galaxy colour and the one determined from the colour-magnitude (CM) relation at a given $F814W$ magnitude. We fit the CM sequence for the spectroscopically confirmed galaxy members by using a robust linear regression \citep{cappellari2013} that considers a possible intrinsic data scatter and clips out outliers, adopting the least trimmed squares technique \citep{Rousseeuw2006}. With this correction, red galaxies are centred around zero, while blue galaxies have colours $\lesssim-0.2$\,mag, regardless of their redshift.

The number of false positives correlates both with the non-GGSL magnitudes and colours for $F814W>19$ and $(F606W-F814W)_{\text{norm}}>-0.5$ (see the approximately constant FPR in the bottom panels of Fig.~\ref{fig:alx_vgg:FP}). There are two FP excesses in the brighter and bluer part of the parameter space. The FP increase for progressively bluer objects ($7\%$ for objects bluer than $-0.5$\,mag, up to $90\%$ and $50\%$, respectively for VGG and SC-VGG, in the bin around $-0.7$\,mag ). These are disc galaxies with a red bulge surrounded by blue spiral-like structures (see the first row in Fig.~\ref{fig:FP:imgs}) or generally blue galaxies, which are under-represented in the KB, since our sample is extracted from cluster cores mainly populated by red members.

The two models also have similar trends as a function of the $F814W$ magnitude, with a constant FPR of $\sim0.16$ for $F814W>19.5$ and an FP excess in the brightest bins. This can be due to embedded lensed features in the training cutouts, which are outshone by the galaxy halo, so that when bright lens galaxies are present, the networks are trained to predict the existence of a GGSL with hidden lensed features (see the second row in Fig.~\ref{fig:FP:imgs}). Indeed, these non-GGSL images appear similar to mock GGSLs with hidden lensed features included in the training set. A bidimensional representation of the distribution of FPRs in the colour-magnitude space is also shown in the top panels of Fig.~\ref{fig:FPFN:2Dhist}, for the VGG and SC-VGG models, which illustrate the trends discussed above. 

\begin{figure}[tb]
   \centering
   \includegraphics[width=0.98\linewidth]{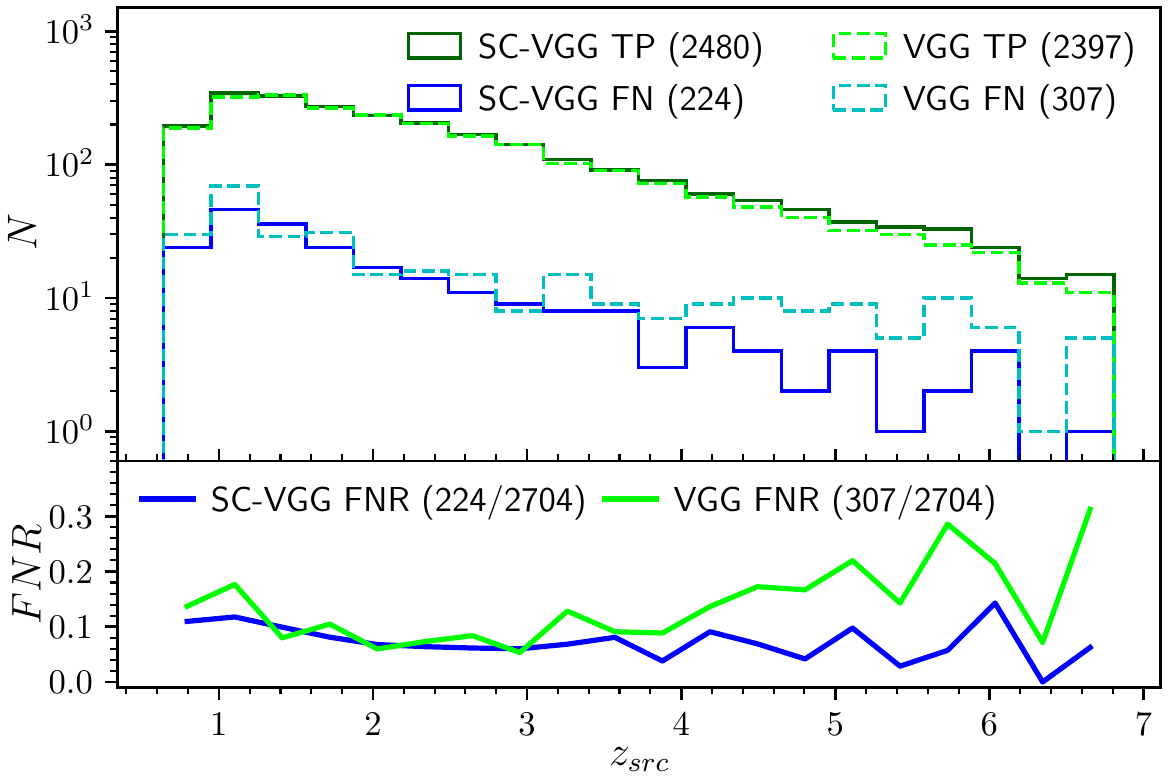}\\\smallskip
   \includegraphics[width=0.98\linewidth]{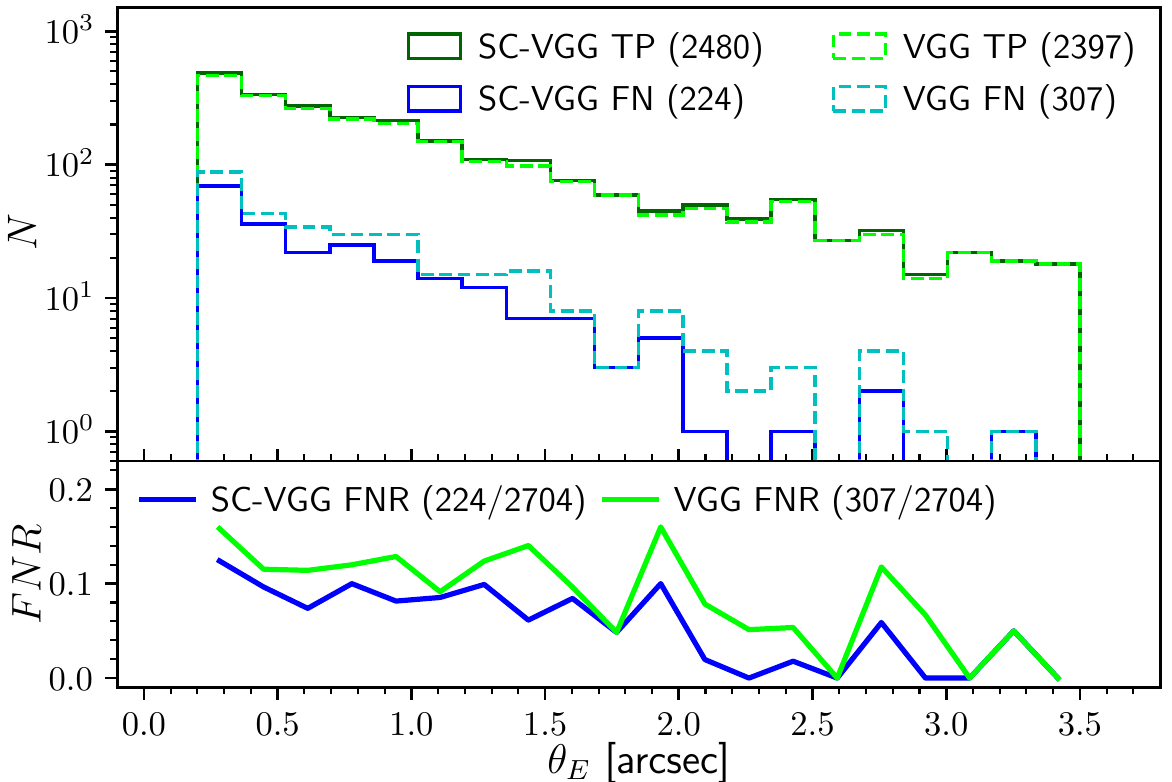}\\\smallskip
   \includegraphics[width=0.98\linewidth]{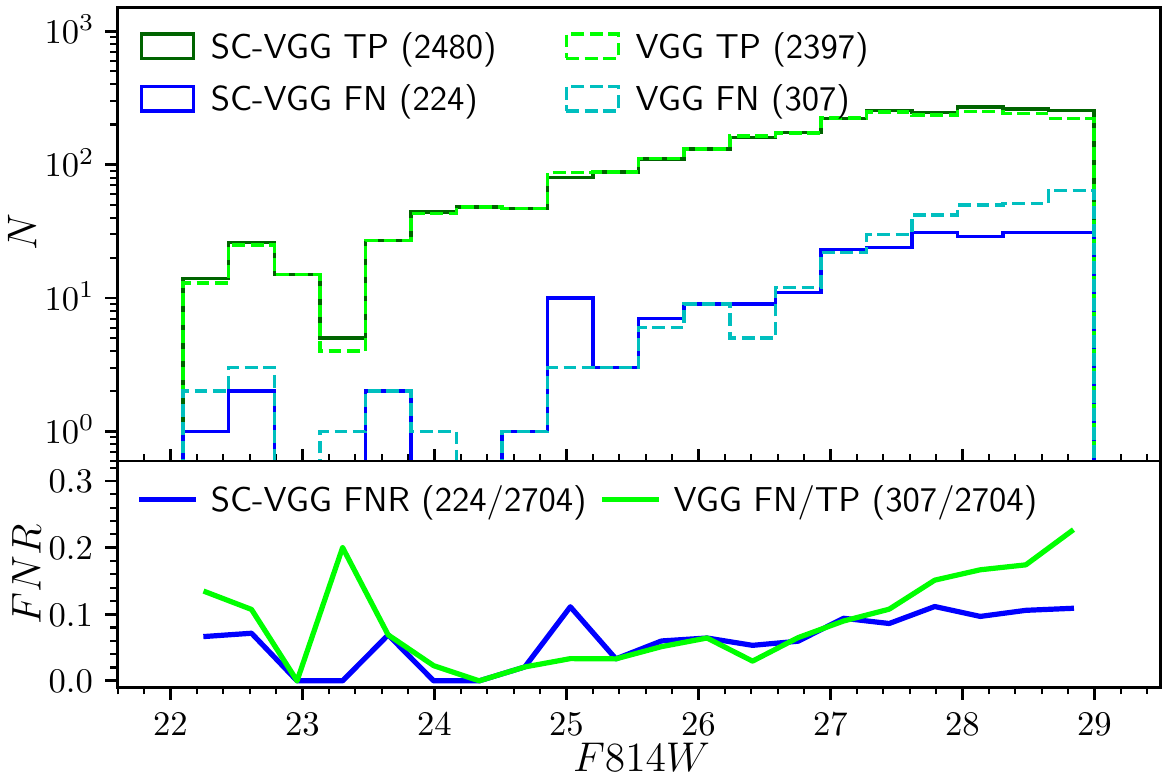}
   \caption{True positive (TP, green lines) and false negative (FP, blue lines) distributions as a function of source redshift ($z_{src}$, top panel), Einstein radius ($\theta_E$, middle panel), and source intrinsic $F814W$ magnitude (bottom panel) for the VGG (dashed lines) and SC-VGG (solid lines) models. The corresponding FN ratio is plotted at the bottom of each panel. In all panels, only sources with available and reliable magnitudes are plotted.}\label{fig:alx_vgg:FN}
\end{figure}

\begin{figure*}[tb]
   \centering
   \includegraphics[width=0.95\linewidth]{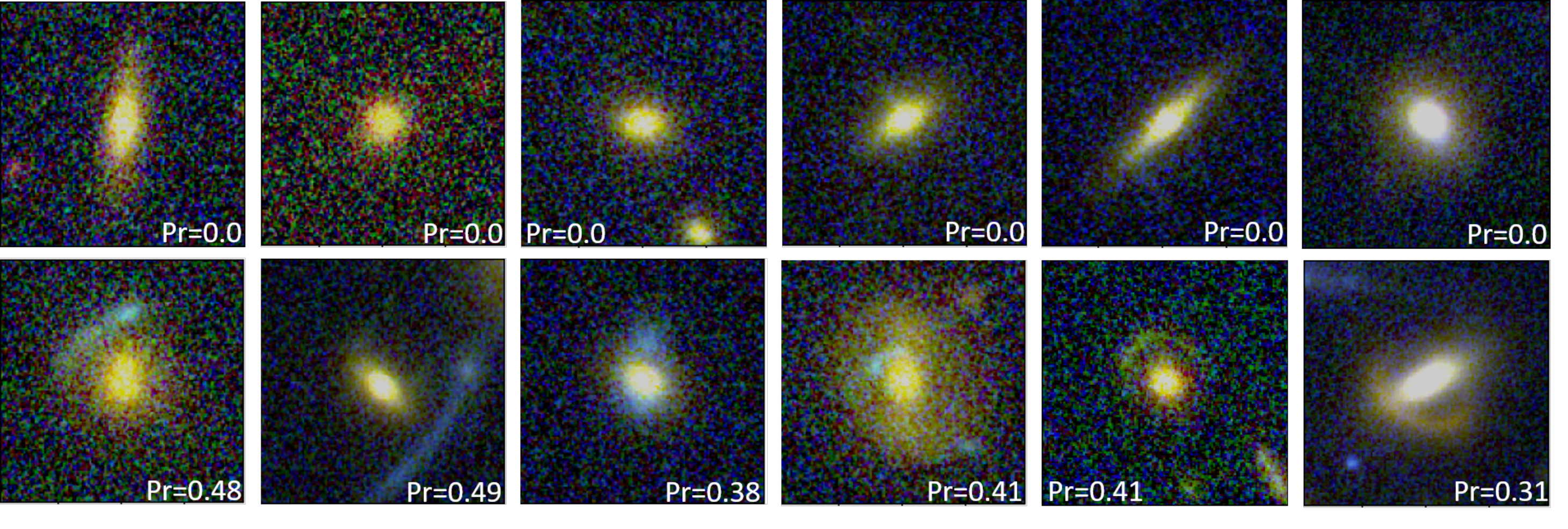}
   \caption{Selection of false negative common to both the VGG and SC-VGG models. The probability of belonging to the GGSL class is shown in each thumbnail (referred to the SC-VGG model). Cutouts are $\sim4\arcsec$ across.}\label{fig:FN:imgs}
\end{figure*}

Regarding the strong-lenses misclassified as non-GGSLs (i.e. the FNs), Fig.~\ref{fig:alx_vgg:FN} shows the distributions of FN, TP and the false negative ratio ($FNR=\frac{FN}{TP+FN}$) as a function of the source redshift (top left panel), galaxy-lens $\theta_E$ (top right panel), and the source intrinsic $F814W$ magnitude (middle right panel). The main dependencies are also summarised in Tab.~\ref{tab:fn_dep}. 
The number of FN decreases with $\theta_E$, with an FNR ratio $\lesssim0.06$ for $\theta_E\gtrsim2\arcsec$. On the other hand, FNs are mainly associated with small-scale galaxy-lenses (FNR $\gtrsim0.10$ for $\theta_E<0.5\arcsec$). Likewise, misclassifications increase with the source magnitude, with an FN fraction $\sim0.10$ for $F814W\ge27$. The VGG and SC-VGG FN ratios are similar ($\sim0.05$) down to $F814W=27$, whereas the VGG FNR continues to increase up to $0.20$ at fainter magnitudes. A 2D-distribution of the FN ratio as a function of $\theta_E$ and $F814W$ is shown in the middle panels of Fig.~\ref{fig:FPFN:2Dhist} for both CNNs. These plots clearly show that the SC-VGG model outperforms the VGG network for faint sources and small-scale lenses (FNR$_{\text{VGG}}\simeq2\times$FNR$_{\text{SC-VGG}}$).

Similarly, the dependence of the FNR on the source redshift for the VGG and SC-VGG models are comparable ($\sim0.10$) up to $z\sim3$, which represents the $70\%$ of the whole FN set. However, the VGG FNR significantly deteriorates (up to $\sim0.21$) at larger redshifts, whereas SC-VGG FNR remains approximately constant. The better performance of the SC-VGG model over the VGG at $z\gtrsim3$ is likely connected to the \textit{drop out} effect for lensed galaxies due to the Lyman-break shift out of the bluest filter. In the VGG model, images in the three filters are combined in the first convolution layer, thus mixing the multi-band information. Instead, with the single-channel approach, only filters which carry information (useful to disentangle GGSLs from non-GGSLs) contribute to the classification. In contrast, the drop-out images (no signal) are down-weighted by the network model. 

We also verify that there is no significant dependence of the FNR on the source effective radius ($r_e$), as illustrated in the bottom panel of Fig.~\ref{fig:FPFN:2Dhist}, where FNR values $\gtrsim0.3$ are mainly confined in the high redshift bin for the VGG model only, as discussed above.

\begin{figure}[tb]
   \centering
   \includegraphics[width=0.95\linewidth]{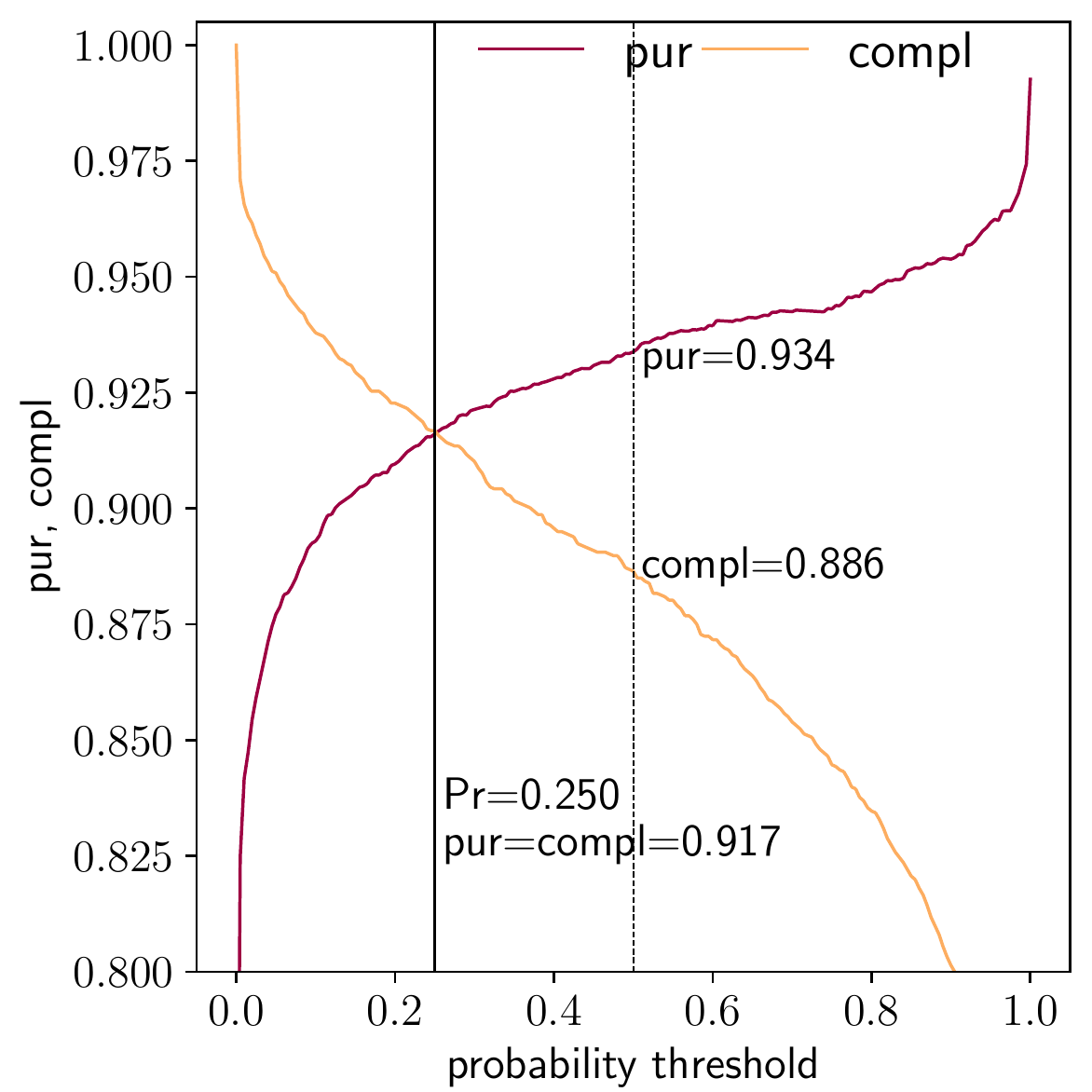}
   \caption{Purity (red) and completeness (orange) as a function of the GGSL probability threshold. Vertical lines correspond to the purity-completeness intersection at $Pr=0.25$ (as a solid line) and to the classical threshold at $Pr=0.50$ (as a dotted line). Purity and completeness values at $Pr=0.25$ and $Pr=0.5$ are shown in the panel.}\label{fig:GGSL:Probs}
\end{figure}

A selection of false negatives is shown in Fig.~\ref{fig:FN:imgs}. The first row includes the adversarial examples containing faint sources ($F814W>27.5$) and small-scale lenses ($0.10\arcsec<\theta_E<0.25\arcsec$). These cases should be compared with the FPs discussed above (see the second row in Fig.~\ref{fig:FP:imgs})  The second row includes bonafide GGSLs, with visible arc-like features. All the adversarial FNs have probabilities (to be a GGSL) equal to zero, whereas bonafide GGSLs have probabilities not far below the adopted threshold of $0.50$. This suggests that these FNs could be recovered by lowering the GGSL probability threshold. For example, one could adopt a threshold corresponding to the value where the purity and the completeness functions intersect. Fig.~\ref{fig:GGSL:Probs} shows that such a value is $Pr=0.25$ for the SC-VGG model. By adopting such a threshold, all FNs in the second row of Fig.~\ref{fig:FN:imgs} are recovered; while analysing the entire sample, we find that the completeness increases by $3.5\%$ at the expense of a $1.8\%$ drop in purity. The optimal strategy on the $Pr$ value will depend on the number of GGSL candidates in a given imaging dataset and specific science objectives.

The analysis carried out in this section can be compared with other studies in blank field (i.e. not in clusters) based on different imaging datasets, adopting a similar methodology. For example, \cite{Petrillo2019} and \cite{Gentile2022} use the KiDS \citep{KiDS2015} and VOICE \citep{Vaccari2016} imaging surveys to search for GGSLs with similar CNNs. They trained their networks with simulated lensed galaxies, adopting simple Singular Isothermal Sphere models for lens early-type galaxies. 
A comparison of our FN distributions as a function of $\theta_E$ (middle panel in Fig.~\ref{fig:alx_vgg:FN}) with those from these studies shows similar FNR ranges, especially in the case of \cite{Petrillo2019}, who found FNR values in the $4\%$--$15\%$ range (see their Fig.~3), while \cite{Gentile2022} obtained larger FNR values ($10\%$--$35\%$, see their Fig.~6).

\section{Searching for strong-lenses in galaxy clusters}\label{sec:run}
The experiments described in the previous sections are mostly focused on the classification efficiency of the image-based CNN with simulated lenses by evaluating its dependence on several observational parameters, such as the magnitude, colour and Einstein radius. In this section, we are mainly interested in evaluating the degree of generalisation achieved by the trained CNNs in classifying real sources as GGSLs. This process step is commonly referred to as \textit{run} in the machine learning context. To maximise the parameter space sampling, we do not use the k-fold approach utilised for performance testing but rather exploit the whole KB by excluding the validation set used for the regularisation processes. 

\begin{figure*}[tb]
   \centering
   \includegraphics[width=0.98\linewidth]{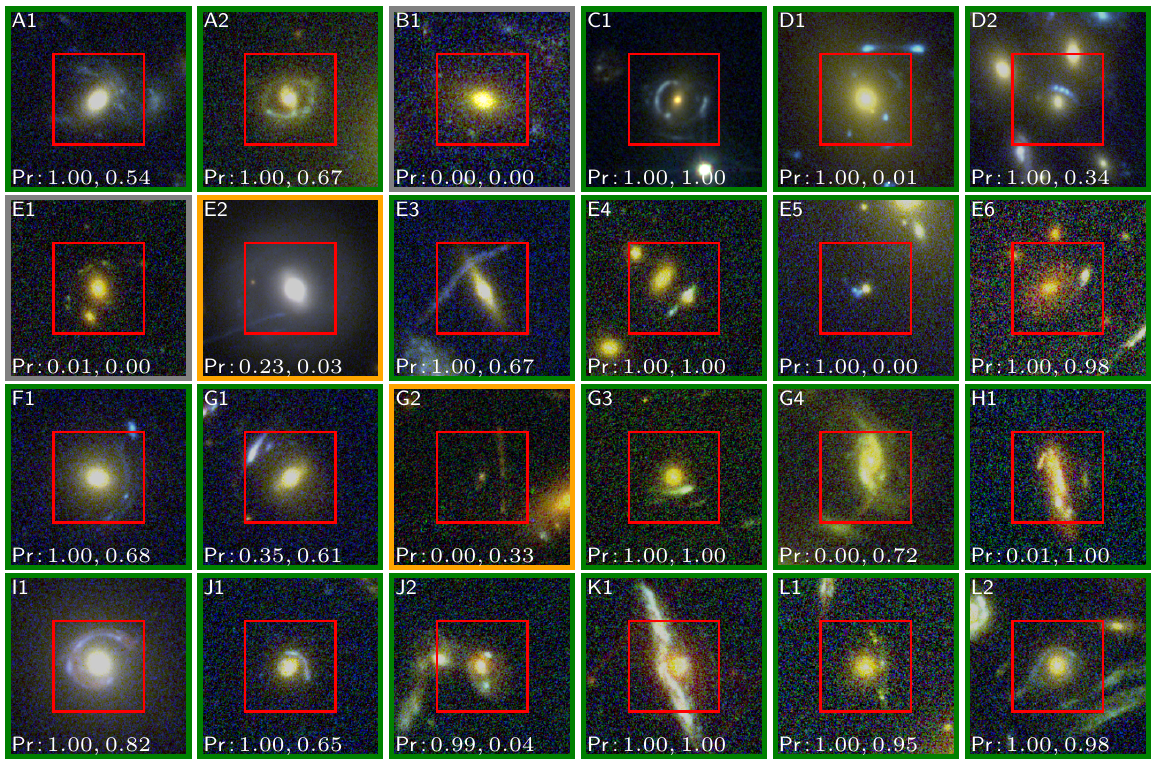}
   \caption{Known GGSLs processed by both VGG and SC-VGG networks (see  Tab.~\ref{tab:GGSL:known}). The GGSL probability is reported in each thumbnail (referred to the VGG and SC-VGG, respectively). Cutouts are $7.7\arcsec$ across. The inner red squares enclose the area actually processed by the networks ($\sim4\arcsec$). According to the classification probability, cutouts are surrounded by a box coloured in green (at least one probability is $>0.5$), orange (at least one probability is $\in (0.2, 0.5]$), and grey (otherwise).}\label{fig:GGSL:known}
\end{figure*}

To test the network generalisation capability, we perform a \textit{run} on $24$ candidate GGSLs previously known in the HST sample of galaxy clusters. These systems are listed in Tab.~\ref{tab:GGSL:known} and shown in Fig.~\ref{fig:GGSL:known}. We assume as ``secure'' those objects whose GGSL nature is based on the presence of clear strong-lensing features, in some cases with a spectroscopic confirmation; those with uncertain classification are flagged as ``uncertain'', i.e. to be confirmed with further observations. We note that some of these secure GGSLs (panels E2, E4, E5, E6, G2) are part of multiple image systems produced by the cluster deflection field in addition to the lens galaxy. We consider $20$ of these candidates as secure. We organize the CNN predictions according to three probability intervals: $Pr>0.5$, $0.2\le Pr <0.5$ and $Pr\le0.2$, respectively defined as true positive (TP), quasi-true positive (qTP) and false negative (FN). Out of the $24$ processed GGSLs, both CNN models yield the same classification for $15$ objects ($17$ by including qTPs), $13$ of which are correctly classified ($15$ by including qTPs). By adopting a probability threshold $Pr>0.2$, out of $20$ secure GGSLs, the true positives are $18$ and $16$ for the VGG and SC-VGG, respectively ($16$ and $14$ by excluding the qTPs).

All typical lenses, with arc-like or ring-like features, have been correctly classified (see, for example, the Einstein rings shown in panels A2, C1, I1 and L2, or the arc-like structures in panels D2, E3, E4, F1, K1 and L1). The system in panel B1 in Fig.~\ref{fig:GGSL:known}, visually identified as a GGSL by \cite{Desprez2018} in Abell~383, is predicted as a non-GGSL by both CNNs. However, a further inspection, including also the HST/WFC3 bands, shows that the faint sources, in a seemingly Einsten cross configuration, have different colors suggesting a correct CNN classification (for this reason, we set this classification as true negative (TN) in Tab.~\ref{tab:GGSL:known}). 

Concerning the FNs, the system in panel E1 is a spectroscopic multiply imaged system in M0416 (named ID.14) studied by \cite{caminha2016} and \cite{Vanzella2017}. This somewhat surprising misclassification may be due to a peculiar lens configuration with two cluster galaxies, which is not well represented in the training set (less than $0.01\%$ of the input lenses). Other FNs (E2, G2) seem to be associated with cluster-scale lensing features, a category which somewhat bridges GGSLs to giant arcs. 
Interestingly, the remaining systems (G4, H1 and J2) with uncertain classification (see Tab.~\ref{tab:GGSL:known}) are characterised by almost complementary probabilities by the two CNN models. This behaviour underscores the hard challenge of the (human or machine-based) GGSL classification process in the presence of peculiar or complex morphologies. 

\section{Conclusions}\label{sec:conclusions}
In this work, we build the methodology to search for galaxy-scale strong lensing systems in the HST multi-band imaging of galaxy clusters by utilising deep learning techniques. We present a novel approach to simulate GGSLs in galaxy clusters, which takes advantage of extensive spectroscopic information on member galaxies in eight clusters whose inner total mass distribution is determined with high accuracy through strong lensing modelling. The accurate knowledge of the deflection field in each cluster allows us to inject background sources near the secondary caustics associated with the cluster galaxies and to simulate highly realistic GGSL systems into the HST cluster field. To this aim, we sample the magnitude and photometric redshift distributions of background galaxies, using \sersic light profiles with a physical size estimated from an empirical redshift evolution of the effective radius of distant galaxies and a given star-forming SED. 

In this way, we generate thousands of mock GGSLs which reproduce the observations with high fidelity, preserving the full complexity of the real data. We use the image cutouts of these simulated GGSL systems in three ACS filters as a knowledge base to train two main CNNs. Their efficiency in identifying and classifying GGSLs in HST images down to $F814W=29$ is quantified using several standard metrics.
The main results of our study can be summarised as follows.
\begin{itemize}
     \item[-] We investigate two CNN architectures: one combines the $F435W$, $F606W$ and $F814W$ ACS bands (VGG model), while the other processes the three channels independently (SC-VGG model). We find that both models achieve a very good trade-off between purity and completeness ($85\%$--$95\%$). This reflects the comprehensive sampling of the parameter space describing the source and lens properties and a highly pure classification of non-GGSL events based on the visual inspection of lensing experts. We also find that performance fluctuations, measured with independent experiments iteratively using a portion of the dataset (the so-called k-fold approach), are within $2\%$--$4\%$, underlying the robustness of the network efficiency.
    \item[-] The analysis of the false positive and false negative rates shows that FPs are typically spiral or disk galaxies whose structure is sometimes mistaken for lensing features. 
    An interesting category of false positives and false negatives includes, respectively, bright galaxies and small cross-section lenses (small Einstein radii), for which the lens galaxy outshines possible multiple images. Although this category encompasses a significant fraction of misclassification, its inclusion in the KB is important to avoid network overfitting.
    \item[-] Overall, the SC-VGG model performs slightly better than the VGG model, based on all the adopted metrics. This is particularly evident for faint and relatively red lensed sources, for which the single-channel approach seems to better take into account the K-correction effects. 
    \item[-] When testing our CNN models on GGSLs previously known from the literature in 12 CLASH and HFF clusters, both networks are able to identify almost all systems deemed as secure GGSLs, thus demonstrating a high degree of generalisation. These true positive cases include a wide range of galaxy-scale strong lensing configurations, while the false negatives seem to be generally associated with GGSLs whose configuration suggests a significant contribution from cluster-scale lensing. 
\end{itemize}

\noindent

 In a forthcoming paper, we plan to perform a systematic search for GGSLs around cluster members in $\sim50$ galaxy clusters, part of several HST programs (CLASH, HFF and RELICS).
In future works, we also intend to extend this methodology to the forthcoming ground and space-based datasets, such as the Euclid \citep{Laureijs2011} and Vera Rubin Observatory \citep{LSST2019} wide-area surveys, as well as the James Webb Space Telescope \citep{Gardner2006} NIRCAM imaging data, whose extraordinary potential in the study of strongly lensed sources has been shown in the first observations of galaxy cluster cores \citep[e.g.][]{Treu2022, Adams2023}.
Moreover, we will explore other deep learning networks, such as deep auto-encoders \citep{goodfellow:2010} and generative adversarial networks \citep{Mirza2014}, in the effort to automate the search and classification of strong-lensing events in these next-generation datasets. In this context, other deep architectures (e.g. region-based CNN, \citealt{ren2015}, or masked region CNN, \citealt{He2017}) can be tested by exploiting the trained convolutional layers developed in this paper.


\begin{acknowledgements}
We thank the anonymous referee for the helpful feedback and suggestions.
We acknowledge financial support through grants PRIN-MIUR 2015W7KAWC, 2017WSCC32, and 2020SKSTHZ.
MB acknowledges financial contributions from the agreement ASI/INAF 2018-23-HH.0, Euclid ESA mission– Phase D.
AA has received funding from the European Union’s Horizon 2020 research and innovation programme under the Marie Skłodowska-Curie grant agreement No 101024195 — ROSEAU. 
MM thanks INAF for support through Minigrant ``The Big-Data era of cluster lensing''. We gratefully acknowledge the support of
NVIDIA Corporation, with the donation of the Titan Xp GPUs used for this
research. \\
In this work several public software was used: Topcat \citep{Taylor05}, Astropy \citep{astropy2013, astropy2018}, TensorFlow \citep{tensorflow2015}, Keras \citep{keras2015} and Scikit-Learn \citep{sklearn:2011}.
\end{acknowledgements}

\bibliographystyle{aa}
\bibliography{aanda}

%
%

\appendix

\section{Complementary tables and figures}\label{app:tabfig}
\begin{figure*}[tbp]
   \centering
   \includegraphics[width=0.95\linewidth]{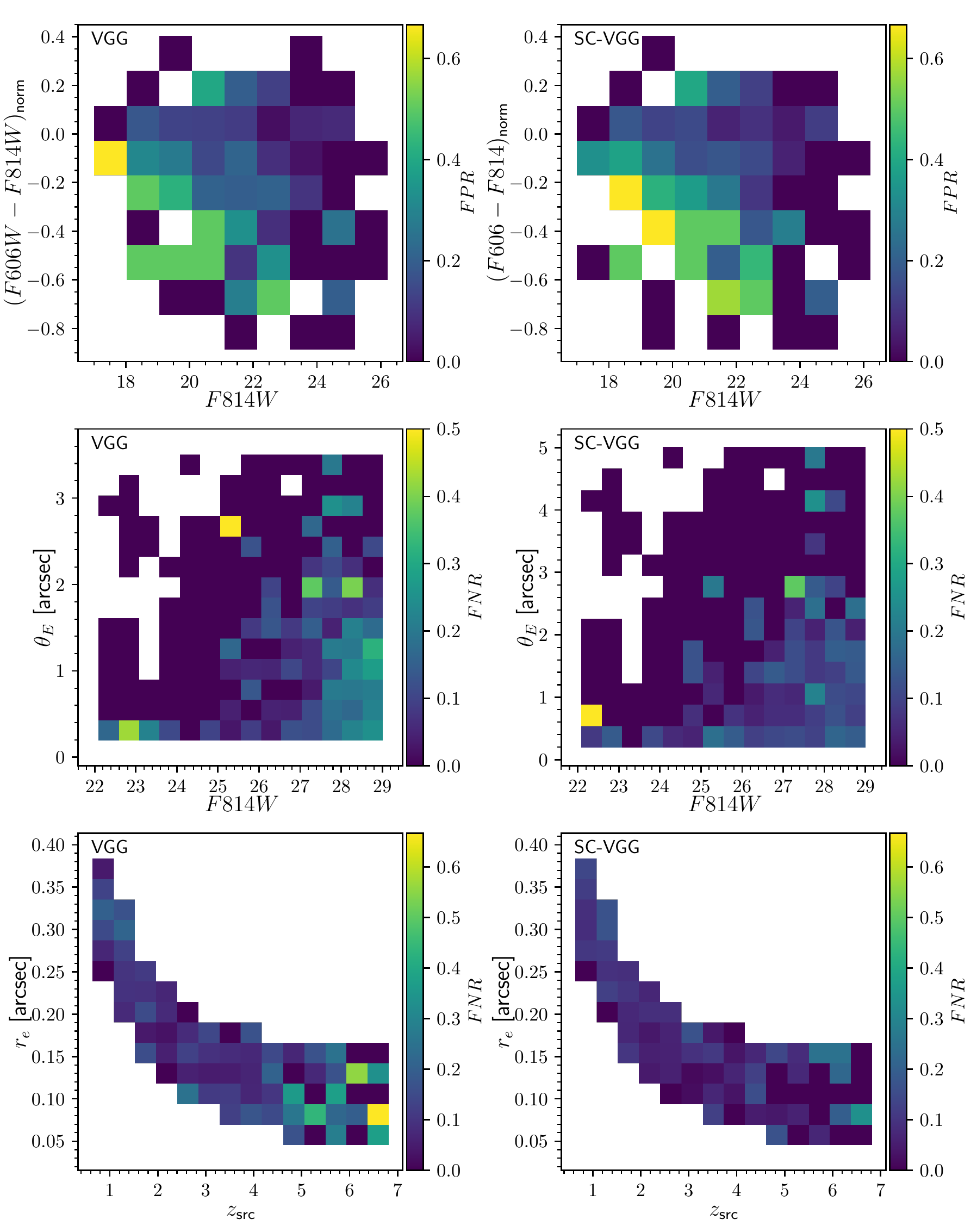}
\caption{False positive and false negative ratio represented as a 2D histogram. \textit{Top panels}: FP ratios on a galaxy colour-magnitude diagram (i.e. $(F606W-F814W)_{\text{norm}}$ vs $F814W$). \textit{Middle panels}: FN ratios on a lens $\theta_E$ vs source $F814W$ magnitude diagram. \textit{Bottom panels}: FN ratios on a source $r_e$ vs source redshift diagram. The VGG and the SC-VGG results are shown on the left and right panels, respectively. The regions of the parameter space with zero true negative or zero true positive values are left white.}\label{fig:FPFN:2Dhist}
\end{figure*}

In this appendix, we include additional tables and figures related to the CNN performance evaluation. Tab.~\ref{tab:vggs_comp} shows a comparison between the VGG and the SC-VGG in terms of statistical estimators, by also excluding the adversarial examples from the metric computation (quoted with an asterisk). The analysis of performance fluctuations (for both VGG and SC-VGG) evaluated over the $10$ folds is summarised in Tab.~\ref{tab:quantiles}. Tab.~\ref{tab:singleband} shows a comparison of the results achieved by the networks trained with the three ACS bands (the adopted method) with performances obtained using a single band.  
A summary of the false positive and false negative distributions is outlined in Tab.~\ref{tab:fp_dep} and Tab.~\ref{tab:fn_dep}, respectively. Fig.~\ref{fig:FPFN:2Dhist} shows the false positive rates and false negative rates as 2D histograms. Tab.~\ref{tab:GGSL:known} illustrates the \textit{run} performed by both VGG and SC-VGG by processing a set of known GGSLs in a sample of galaxy clusters observed with HST.

\begin{table}[htb]\caption{Performance comparison between the two CNN architectures.} \label{tab:vggs_comp}
\centering
\begin{tabular}{llcccc}
\hline
& [\%] & VGG & VGG* & SC-VGG & SC-VGG* \\\hline
& \textit{AE}          & 87.7 & 89.6 & 89.4 & 89.5 \\
\noalign{\vskip 2mm}
\multirow{3}{*}{\rotatebox[origin=c]{90}{GGSL}}
& \textit{pur}         & 93.4 & 87.5 & 93.1 & 86.7   \\
& \textit{compl}  & 88.6 & 93.4 & 91.7 & 94.5   \\
& \textit{F1}          & 91.0 & 90.4 & 92.4 & 90.4   \\
\noalign{\vskip 2mm}
\multirow{3}{*}{\rotatebox[origin=c]{90}{NGGSL}}
& \textit{pur}         & 76.7 & 92.3 & 81.4 & 93.3   \\
& \textit{compl} & 85.4 & 85.4 & 84.1 & 84.1   \\
& \textit{F1}          & 81.1 & 88.7 & 82.8 & 88.5   \\
\noalign{\vskip 2mm}\hline
\end{tabular}
\tablefoot{Network pperformances are reevaluated by removing faint sources and small-scale lenses ($F814W>28$\,mag and $\theta_E<0.5\arcsec$) are marked by an asterisk.}
\end{table}

\begin{table}[htb]\caption{Fluctuations of the performances for the VGG and SC-VGG models.} \label{tab:quantiles}
\resizebox{\linewidth}{!}{
\begin{tabular}{llcccccccccccc}\hline
&               &  \multicolumn{2}{c}{median} & \multicolumn{2}{c}{$Q_1$} & \multicolumn{2}{c}{$Q_3$} \\\hline
& [\%]       & VGG & SC-VGG & VGG & SC-VGG & VGG & SC-VGG \\\hline
& \textit{AE}          & \textbf{88.3} & \textbf{89.3} & 86.2 & \textbf{88.6} & 88.6 & \textbf{90.3} \\
\noalign{\vskip 2mm}
\multirow{3}{*}{\rotatebox[origin=c]{90}{GGSL}} & \textit{pur}         & \textbf{93.3} & \textbf{93.3} & \textbf{92.5} & 92.3 & \textbf{94.8} & 94.2 \\
& \textit{compl}  & 89.1 & \textbf{91.9} & 86.7 & \textbf{91.2} & 90.4 & \textbf{92.5} \\
& \textit{F1}          & 91.5 & \textbf{92.3} & 89.8 & \textbf{91.8} & 91.5 & \textbf{93.0} \\
\noalign{\vskip 2mm}
\multirow{3}{*}{\rotatebox[origin=c]{90}{NGGSL}} & \textit{pur}         & 77.6 & \textbf{81.9} & 74.0 & \textbf{79.7} & 79.0 & \textbf{82.9} \\
& \textit{compl} & \textbf{85.6} & 84.2 & \textbf{82.3} & 81.7 & \textbf{88.1} & 87.2 \\
& \textit{F1}          & 91.5 & \textbf{92.3} & 89.8 & \textbf{91.8} & 91.5 & \textbf{93.0} \\
\noalign{\vskip 2mm} 
\hline

&&\multicolumn{2}{c}{$IQR$} & \multicolumn{2}{c}{$Q_1-1.5\cdot IQR$} & \multicolumn{2}{c}{$Q_3+1.5\cdot IQR$} \\\hline
&[\%]     & VGG & SC-VGG & VGG & SC-VGG & VGG & SC-VGG \\\hline
& \textit{AE}          & 2.4 & \textbf{1.7} & 85.3 & \textbf{87.8} & 89.7 & \textbf{91.4} \\
\noalign{\vskip 2mm}
\multirow{3}{*}{\rotatebox[origin=c]{90}{GGSL}} & \textit{pur} & 2.3 & \textbf{1.9} & \textbf{90.7} & 90.1 & 95.6 & \textbf{95.7} \\
& \textit{compl}  & 3.7 & \textbf{1.4} & 84.1 & \textbf{90.4} & 92.5 & \textbf{94.1} \\
& \textit{F1}          & 1.8 & \textbf{1.2} & 88.8 & \textbf{91.0} & 92.6 & \textbf{93.9} \\
\noalign{\vskip 2mm}
\multirow{3}{*}{\rotatebox[origin=c]{90}{NGGSL}} & \textit{pur}         & 5.0 & \textbf{3.2} & 70.5 & \textbf{76.1} & 81.8 & \textbf{85.6} \\
& \textit{compl} & 5.8 & \textbf{5.4} & \textbf{79.7} & 77.6 & \textbf{90.7} & \textbf{90.7} \\
& \textit{F1}          & 1.8 & \textbf{1.2} & 88.8 & \textbf{91.0} & 92.6 & \textbf{93.9} \\
\noalign{\vskip 2mm}
\hline
\end{tabular}
}
\tablefoot{$Q_1$ and $Q_3$ are the 25th and 75th percentiles. The inter-quartile range $IQR=Q_3-Q_1$; the range $(Q_1-1.5\cdot IQR,\, Q_3+1.5\cdot IQR)$ encloses the metric fluctuation within $\pm2.698\sigma$. The best results are highlighted in bold. Average efficiency and GGSL estimators are graphically shown in the bottom panels of Fig.~\ref{fig:vggs_comp}. }
\end{table}

\begin{table}[tb]\caption{Network performances trained with 3 HST/ACS bands (VGG, SC-VGG) compared with single band training. } \label{tab:singleband}
\centering
\begin{tabular}{llccccc}
\hline
& [\%] & VGG & SC-VGG & $F435W$ & $F606W$ & $F814W$  \\\hline
& \textit{AE}          & 87.7 & 89.4 & 87.2 & 86.1 & 86.8 \\
\noalign{\vskip 2mm}
\multirow{3}{*}{\rotatebox[origin=c]{90}{GGSL}}
& \textit{pur}         & 93.4 & 93.1 & 91.8 & 91.1 & 91.5 \\
& \textit{compl}  & 88.6 & 91.7 & 89.8 & 88.8 & 89.3 \\
& \textit{F1}          & 91.0 & 92.4 & 90.8 & 89.9 & 90.4 \\
\noalign{\vskip 2mm}
\multirow{3}{*}{\rotatebox[origin=c]{90}{NGGSL}}
& \textit{pur}         & 76.7 & 81.4 & 77.5 & 75.4 & 76.6 \\
& \textit{compl} & 85.4 & 84.1 & 81.3 & 79.9 & 80.9 \\
& \textit{F1}          & 81.1 & 82.8 & 79.3 & 77.6 & 78.7 \\
\noalign{\vskip 2mm}
\hline
\end{tabular}
\end{table}

\begin{table}[tb]\caption{Summary of false positive distributions for the VGG and SC-VGG networks.} \label{tab:fp_dep}
\resizebox{\linewidth}{!}{
\begin{tabular}{lccccc}\hline
&& \multicolumn{2}{c}{VGG} &  \multicolumn{2}{c}{SC-VGG} \\\hline
                                     & \bf{NGGSL} & FP & FP/TN & FP & FP/TN \\\hline
Total Number     & 1037   & 154    & 0.174 & 170    & 0.196\\\hline
\noalign{\vskip 2mm}
$F814W<19.5$      & 9.6\%  & 16.8\% & 0.263 & 15.3\% & 0.356\\ 
$F814W\ge19.5$    & 90.4\% & 83.2\% & 0.158 & 84.7\% & 0.181\\ \hline
\noalign{\vskip 2mm}
colour\,$<-0.5$ & 3.9\%  & 6.5\%  & 0.333 & 7.6\%  & 0.481\\
colour\,$\ge-0.5$   & 96.1\% & 93.5\% & 0.169 & 92.3\% & 0.187\\\hline
\end{tabular} }
\tablefoot{Fractions of NGGSL (Col.~2), FP (Col.~3 and Col.~5) and FN to TN ratio (Col.~4 and Col.~6) as a function of source magnitude (second and third row) and galaxy normalized colour (i.e. $(F606W-F814W)_{\text{norm}}$, fourth and fifth row). The total number of spectroscopic NGGLSs and FPs are quoted in the first row.}
\end{table}

\begin{table}[tb]\caption{Summary of the false negative distributions, split between VGG and SC-VGG network.} \label{tab:fn_dep}
\resizebox{\linewidth}{!}{
\begin{tabular}{lccccc}\hline
&& \multicolumn{2}{c}{VGG} & \multicolumn{2}{c}{SC-VGG} \\\hline
                                     & \bf{GGSL} & FN & FN/TP & FN & FN/TP \\\hline
Total Number                         & 2704   & 307    & 0.128 & 224    & 0.090\\\hline
\noalign{\vskip 2mm}
$F814W\ge28.0$                        & 31.1\% & 52.1\% & 0.235 & 38.8\% & 0.115\\ 
$F814W\ge27.0$                        & 61.0\% & 83.1\% & 0.183 & 73.2\% & 0.110\\ 
$F814W<27.0$                          & 39.0\% & 17.9\% & 0.052 & 26.8\% & 0.060\\\hline
\noalign{\vskip 2mm}
$\theta_E<0.5\arcsec$                & 32.2\% & 41.4\% & 0.171 & 46.9\% & 0.137\\
$\theta_E\ge0.5\arcsec$              & 67.8\% & 58.6\% & 0.109 & 53.1\% & 0.069\\\hline
\noalign{\vskip 2mm}
$z_{src}\ge5$                        & 5.9\%  & 11.4\% & 0.282 & 4.9\%  & 0.074\\
$z_{src}\ge4$                        & 12.9\% & 20.8\% & 0.225 & 10.7\% & 0.074\\
$z_{src}\ge3$                        & 25.5\% & 31.9\% & 0.166 & 20.1\% & 0.070\\
$z_{src}<3$                          & 74.5\% & 68.1\% & 0.116 & 70.9\% & 0.098\\\hline

\end{tabular} }
\tablefoot{Fractions of GGSL (Col.~2), FN (Col.~3 and Col.~5) and FN to TP ratio (Col.~4 and Col.~6) as a function of source magnitude (second to fourth row), lens galaxy $\theta_E$ (fifth and sixth row) and source redshift (seventh to eighth row). The total number of GGLSs and FNs are quoted in the first row.}
\end{table}

\begin{table*}[tb]\caption{Catalogue of known GGSLs processed by both the VGG and SC-VGG networks}\label{tab:GGSL:known}
\centering
\begin{tabular}{cccccccl}
\hline
Cluster & RA & DEC & Image & VGG & SC-VGG & ref &\\\hline
A209  & 22.95776  & -13.60326 & A1 & TP & TP & (1) & secure \\
A209  & 22.96488  & -13.63631 & A2 & TP & TP & (1) & secure\\
A383  & 42.01136  & -3.54803  & B1 & TN & TN & (1) & no\\
M0329 & 52.42013  & -2.22163  & C1 & TP & TP & (1) & secure, $z=1.112$\\
M1149 & 177.40389 & 22.42663  & D1 & TP & FN & (2) & secure, $z=1.806$\\
M1149 & 177.39314 & 22.41134  & D2 & TP & qTP& (2) & secure\\
M0416 & 64.03408  & -24.06675 & E1 & FN & FN & (3) & secure, $z=3.222$\\
M0416 & 64.02847  & -24.08567 & E2 & qTP& FN & (1,5) & secure, $z=2.218$\\
M0416 & 64.01709  & -24.08955 & E3 & TP & TP & (4) & secure\\
M0416 & 64.03262  & -24.06838 & E4 & TP & TP & (5) & secure, $z=2.095$\\
M0416 & 64.03250  & -24.07849 & E5 & TP & FN & (5) & secure, $z=2.542$\\
M0416 & 64.02442  & -24.08106 & E6 & TP & TP & (5) & secure, $z=1.964$\\
M1115 & 168.95626 & 1.49741   & F1 & TP & TP & (1) & secure\\
R2248 & 342.15574 & -44.54591 & G1 & qTP & TP & (1) & secure, $z=0.9406$\\
R2248 & 342.16336 & -44.52972 & G2 & FN & qTP & (1) & secure\\
R2248 & 342.18205 & -44.54035 & G3 & TP & TP & (6) & secure, $z=1.837$ \\
R2248 & 342.17554 & -44.53558 & G4 & FN & TP & (6) & uncertain\\
R1347 & 206.89603 & -11.75360 & H1 & FN & TP & (1) & uncertain\\
R2129 & 22.42878  & 0.10807   & I1 & TP & TP & (1) & secure\\
M0429 & 67.40208  & -2.87139  & J1 & TP & TP & (1) & secure\\
M0429 & 67.38925  & -2.87412  & J2 & TP & FN & (1) & uncertain\\
M0744 & 116.21217 & 39.45987  & K1 & TP & TP &  (1) & secure\\
M1206 & 181.56667 & -08.80478 & L1 & TP & TP & (7) & secure, $z=3.752$\\
M1206 & 181.55309 & -08.79486 & L2 & TP & TP & (7) & secure, $z=1.425$\\\hline
&&         & N$_{\text{TP}}$   & 17  & 14 & \\
&& TOTAL   & N$_{\text{qTP}}$  &  2  &  2 & \\
&&         & N$_{\text{FN}}$   &  4  &  5 & \\\hline

\end{tabular}
\tablefoot{Based on GGSL probability computed by the two CNN models, systems are classified as: true positive (TP, $Pr>0.5$), quasi-true Positive (qTP, $0.2\le Pr <0.5$), false negative (FN, $Pr<0.2$). The last column refers to  the classification as GGSL: ``secure'' for bonafide galaxy-scale systems (with the source redshift when available), ``uncertain'' for those which need verification and ``no'' for non-GGSL systems. See Sec.~\ref{sec:run} for details. The total number of TPs, qTPs and FNs, shown at the bottom of the table, are computed by considering only the secure systems.      }
\tablebib{(1)~\cite{Desprez2018}; (2)~\cite{Smith2005}; (3)~\cite{Vanzella2017};
(4)~\cite{Diego2015}; (5)~\cite{Bergamini_m0416}; (6)~\cite{caminha2016}; (7)~\cite{bergamini2019}.  }
\end{table*}

\end{document}